%% file: quartz-v9.tex
\documentclass[aps, prb, twocolumn, reprint, superscriptaddress, floatfix, citeautoscript, footinbib, longbibliography]{revtex4-2}

\input{preamble/packages.tex}

\input{preamble/abbreviations.tex}

\input{preamble/commands.tex}
\begin{document} 

\title{Lithium-ion dynamics in synthetic quartz studied via the NMR of implanted \ce{^{8}Li^{+}}}

\author{W.~Andrew~MacFarlane}
\email[Email:]{wam@chem.ubc.ca}
\affiliation{Department of Chemistry, University of British Columbia, Vancouver, BC V6T~1Z1, Canada}
\affiliation{Stewart Blusson Quantum Matter Institute, University of British Columbia, Vancouver, BC V6T~1Z4, Canada}
\affiliation{TRIUMF, Vancouver, BC V6T~2A3, Canada}

\author{Ryan~M.~L.~McFadden}
\altaffiliation[Current address: ]{TRIUMF, Vancouver, BC V6T~2A3, Canada}
\affiliation{Department of Chemistry, University of British Columbia, Vancouver, BC V6T~1Z1, Canada}
\affiliation{Stewart Blusson Quantum Matter Institute, University of British Columbia, Vancouver, BC V6T~1Z4, Canada}

\author{Signy~Spencer}
\affiliation{Department of Chemistry, University of British Columbia, Vancouver, BC V6T~1Z1, Canada}
\affiliation{Stewart Blusson Quantum Matter Institute, University of British Columbia, Vancouver, BC V6T~1Z4, Canada}

\author{Aris~Chatzichristos}
\altaffiliation[Current address: ]{Accenture, Athens 145 64, Greece}
\affiliation{Stewart Blusson Quantum Matter Institute, University of British Columbia, Vancouver, BC V6T~1Z4, Canada}
\affiliation{Department of Physics, University of British Columbia, Vancouver, BC V6T~1Z1, Canada}

\author{John~O.~Ticknor}
\affiliation{Department of Chemistry, University of British Columbia, Vancouver, BC V6T~1Z1, Canada}
\affiliation{Stewart Blusson Quantum Matter Institute, University of British Columbia, Vancouver, BC V6T~1Z4, Canada}

\author{David~L.~Cortie}
\altaffiliation[Current address: ]{Australian Nuclear Science and Technology Organisation, Lucas Heights, New South Wales 2234, Australia}
\affiliation{Stewart Blusson Quantum Matter Institute, University of British Columbia, Vancouver, BC V6T~1Z4, Canada}

\author{Martin~H.~Dehn}
\altaffiliation[Current address: ]{D-Wave Systems, Burnaby, BC V5G~4M9, Canada}
\affiliation{Stewart Blusson Quantum Matter Institute, University of British Columbia, Vancouver, BC V6T~1Z4, Canada}
\affiliation{Department of Physics, University of British Columbia, Vancouver, BC V6T~1Z1, Canada}

\author{Sarah~R.~Dunsiger}
\affiliation{TRIUMF, Vancouver, BC V6T~2A3, Canada}

\author{Derek~Fujimoto}
\altaffiliation[Current address: ]{TRIUMF, Vancouver, BC V6T~2A3, Canada}
\affiliation{Stewart Blusson Quantum Matter Institute, University of British Columbia, Vancouver, BC V6T~1Z4, Canada}
\affiliation{Department of Physics, University of British Columbia, Vancouver, BC V6T~1Z1, Canada}

\author{Z.~H.~Jang}
\affiliation{Department of Physics, Kookmin University, 77 Jeongneung-ro, Seongbuk-gu, Seoul, 02707, Korea}

\author{Victoria~L.~Karner}
\altaffiliation[Current address: ]{TRIUMF, Vancouver, BC V6T~2A3, Canada}
\affiliation{Department of Chemistry, University of British Columbia, Vancouver, BC V6T~1Z1, Canada}
\affiliation{Stewart Blusson Quantum Matter Institute, University of British Columbia, Vancouver, BC V6T~1Z4, Canada}

\author{Robert~F.~Kiefl}
\affiliation{Department of Physics, University of British Columbia, Vancouver, BC V6T~1Z1, Canada}
\affiliation{Stewart Blusson Quantum Matter Institute, University of British Columbia, Vancouver, BC V6T~1Z4, Canada}
\affiliation{TRIUMF, Vancouver, BC V6T~2A3, Canada}

\author{Gerald~D.~Morris}
\affiliation{TRIUMF, Vancouver, BC V6T~2A3, Canada}

\author{Monika~Stachura}
\affiliation{TRIUMF, Vancouver, BC V6T~2A3, Canada}
\affiliation{Department of Chemistry, Simon Fraser University, Burnaby, BC V5A 1S6, Canada}

\date{\today}

\begin{abstract}
We report $\beta$-detected nuclear magnetic resonance (\bnmr) measurements of implanted $^8$Li$^+$ in a synthetic single crystal of $\alpha$-SiO$_2$ (quartz).
At 6.55 Tesla, the spectrum is comprised of a large amplitude broad resonance and a quadrupolar multiplet that
is only revealed by an RF comb excitation. The quadrupole splitting is surprisingly small, increases with temperature,
and provides information on the implantation site. Supercell density functional theory calculations show that
the small EFG is consistent with an in-channel interstitial site (Wyckoff $3a$).
The spin-lattice relaxation is unexpectedly fast and strongly temperature dependent
with a diffusive peak above 200 K and a second more prominent relaxation peak at lower temperature.
Analysis of the diffusive relaxation yields an activation barrier 178(43) meV for the isolated \lip\ in the range of other measurements and calculations.
To account for many of the other features of the data, it is suggested that some of the implanted ions trap
an electron forming the neutral \liz, which is stable over a narrow range of temperatures.
\end{abstract}

\maketitle

\section{Introduction \label{sec:introduction}}

\begin{figure}
\centering        
\includegraphics[width=\columnwidth]{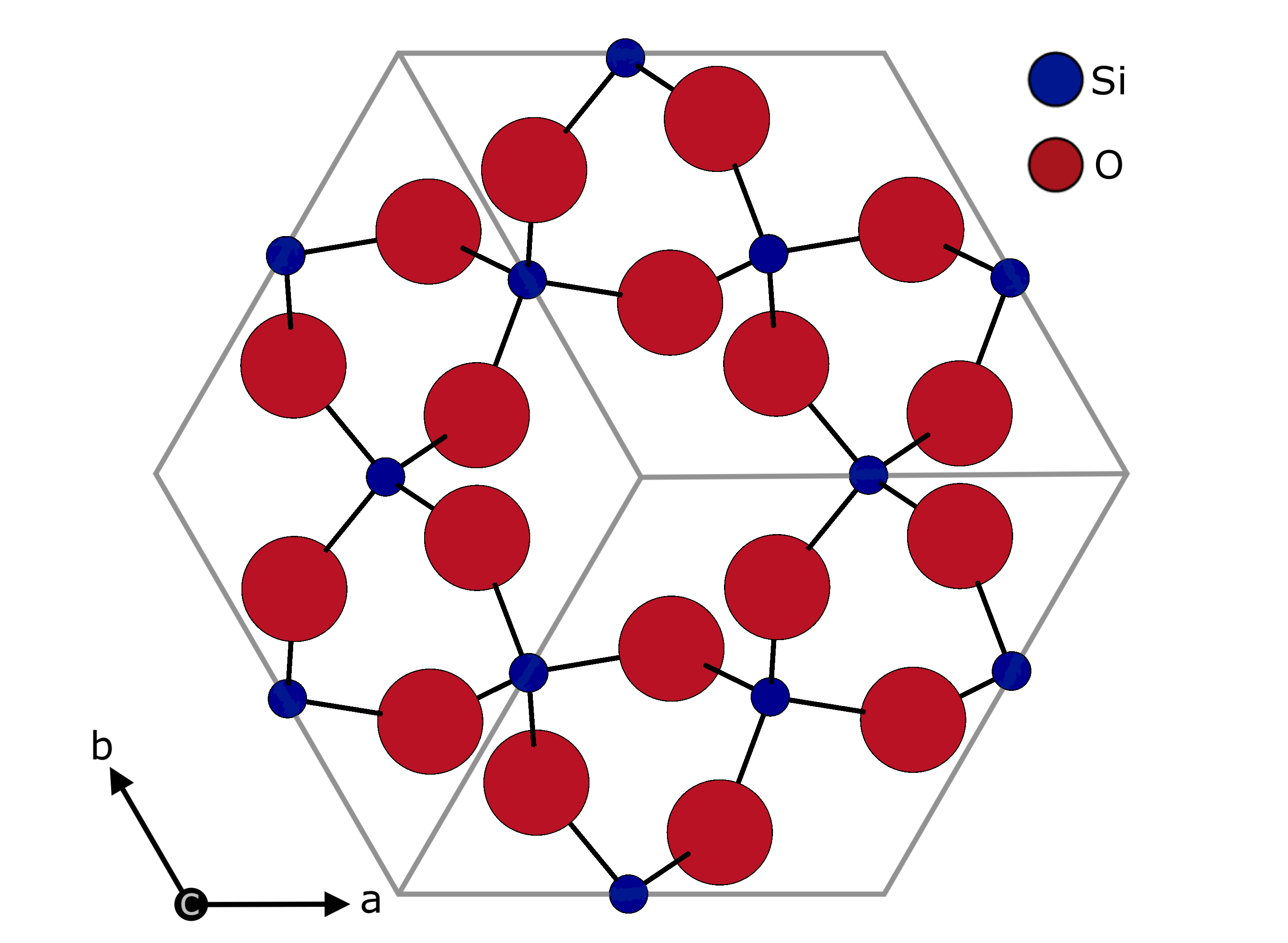} 
\caption{Projection of the structure along the $c$-axis showing the
large open channel centered on the $c$ edge of
the conventional unit cell shown in \Cref{dft}.}
\label{fig:channel}
\end{figure}

Crystalline quartz (SiO$_2$) in its low temperature $\alpha$-polytype is a wide band gap oxide insulator
useful for its piezoelectric, optical and robust high temperature properties.
Its chemical simplicity conceals significant complexity in its crystal structure
composed of chains of corner sharing SiO$_4$ tetrahedra winding along the $c$-axis to form a chiral crystal (space group $P3_121$ or $P3_221$ depending on the handedness\cite{Glazer2018})
as demonstrated, for example, by its optical rotatory power\cite{Einot2006}.
The linked chains of tetrahedra surround open channels as illustrated by the $c$-axis projection shown in \Cref{fig:channel}.
These channels admit highly one-dimensional (1D) mobility of small interstitial ions such as \lip, and this forms the motivation for
the experiments reported here, where we use nuclear magnetic resonance (NMR) to study the short-lived radioisotope \elip\ implanted
into high purity synthetic quartz.
Previously, we had found very interesting dynamic properties of implanted \elip\ in another 1D \lip\ conductor,
rutile TiO$_2$\cite{McFadden2017,Chatzichristos2019},
and we were interested to see if this was more general.
Even though the macroscopic \lip\ conductivity of quartz is quite low\cite{Verhoogen1952,Wenden1957},
there are solid state battery designs featuring \ce{SiO2} in the solid electrolyte interphase (SEI)
layer at a Si electrode\cite{Kushida2002,Chan2009,Kim2014,Zhang2014}.
The properties of interstitial alkalis are also of geological interest in various forms of natural quartz\cite{Gotze2021,Phelps2022}.
Even in synthetic crystals\cite{Brice1985}, they are known to contribute to dielectric loss\cite{SnowGibbs1964,Martin1996} limiting the performance
of some piezoelectric devices and motivating development of electrochemical sweeping methods\cite{MartinSweepRev1988}.

Dimensional confinement significantly modifies diffusive dynamics\cite{Burada2009}. The most extreme case that still allows long range transport
is when the motion is confined to 1D, which is relevant for many real-world systems (e.g., ion channels that traverse the cell membrane,
water in nanoporous media like carbon nanotubes\cite{Chatzichristos2022} or microfluidic devices,  
and atomic diffusion in certain crystal matrices where the structure naturally provides 1D channels such as \quartz).
A distinguishing feature of 1D diffusion is that the mean square displacement of a diffuser grows with time as $\sqrt{t}$ rather than linearly, i.e.\
it is intrinsically ``subdiffusive''\cite{Beijeren1983,Abel2009,Wei2000}.
The 1D character of the motion of \lip\ in quartz is demonstrated by its highly anisotropic conductivity
which is several orders of magnitude larger parallel to the channels, with a temperature dependence governed by a
relatively large energy barrier ($E_a \sim 1$ eV)\cite{Verhoogen1952,Wenden1957}.
However, this $E_a$ is not likely intrinsic. Rather, it probably represents the binding energy at dilute crystal defects
that act as trap sites. For example, Al$^{3+}$ (substitutional for Si$^{4+}$) has a relative charge of $-1$
which binds a small mobile interstitial cation (H$^+$, Li$^+$, Na$^+$) in the adjacent channel.
Trapping at such charged defects is likely the rate limiting process for the macroscopic ionic conductivity.
This is confirmed by radiation induced conductivity (RIC) measurements\cite{Hughes1975},
where holes from electron-hole pair excitations can localize at an Al$^{3+}$ impurity, neutralizing it and releasing the bound alkali\cite{Bahadur2006}.
There is good evidence that the resulting RIC is due to mobility of the liberated alkalis\cite{Jain1982,Martini1986,Miyazaki1989}, 
and it shows a much smaller $E_a \sim 140$ meV, which is probably the intrinsic barrier\cite{Jain1982}.
Remarkably, the trapping effect of Al vanishes in the dense limit in the structurally related
superionic conductor $\beta$-eucryptite \ce{LiSiAlO4}\cite{Follstaedt1976}.
That the liberated \lip\ becomes mobile above $\sim 150$ K is further demonstrated by the production of the neutral \liz\
by a two step (Xray) irradiation. The first step (which frees the \lip\ from an Al$^{3+}$) must be done above 150 K to allow the \lip\ to thermally diffuse some distance
from the Al hole center [AlO$_4$]$^0$\cite{Weil1984}. The sample is then cooled to 77 K (freezing the now isolated interstitial \lip) and a second irradiation step
provides a radiolytic electron that is captured by the \lip\ to form the neutral.

Here, we use the NMR spin-lattice relaxation (SLR) rate $1/T_1$ to sense diffusive dynamics.
This is the rate at which the initially polarized nuclear spin relaxes towards thermal equilibrium where the polarization is nearly zero.
It is determined by fluctuations of the local electromagnetic field at the NMR frequency $\omega_0$ that, for a diffusing species, reflect the stochastic site-to-site hopping.
This quantity has been used extensively to study diffusion in solids at the microscopic scale\cite{BoyceHuberman1979,Richards1979,Kanert1982,Brinkmann1992,Barnes1997,Bohmer2007,Heitjans2012,Habasaki2017}.
Its primary feature is a maximum in  $1/T_1(T)$ when the fluctuation rate (the elementary hop rate $1/\tau$) matches $\omega_0$, known as
the Bloembergen-Purcell-Pound (BPP) peak\cite{BPP1948}. Based on the RIC,
we anticipate the BPP peak will be above $\sim 150$ K.
The strong $T$ dependence of the rate on either side of the BPP peak reflects the activated $T$ dependence of the hop rate, 
\begin{equation}
\frac{1}{\tau} = \frac{1}{\tau_0} e^{-\frac{E_a}{k_{B}T}},
\label{eq:arr}
\end{equation}
where $E_a$ is the barrier to hopping, $k_{B}$ is the Boltzmann constant, and $1/\tau_0$ is the exponential prefactor.
In an elementary microscopic picture, $1/\tau_0$ represents a typical vibrational frequency
on the order of $10^{12}$ Hz. The fluctuation spectral density for stochastic hopping depends on dimensionality \cite{Richards1979,Heitjans2012} resulting
in a specific $\omega_0$ and $T$ dependence of $1/T_1$. However, experimentally in solids with very high ionic mobility, there are often
deviations from this ideal behavior. For example, the prefactor extracted from $1/T_1(T)$ may differ by several orders of magnitude from its expected range.
Such prefactor anomalies\cite{Villa1983,BoyceHuberman1979} may be influenced by both dimensionality and disorder, though they remain poorly understood\cite{Chowdhury2014}.
The anomalous prefactors we found in TiO$_2$\cite{McFadden2017} suggested that we may be able to study this phenomenon more systematically by implanting the same
mobile \elip\ into a number of different 1D hosts in the ultra-dilute limit.

In $\alpha$-SiO$_2$, we find the spin-lattice relaxation for \eli\ is unexpectedly fast and strongly temperature dependent
with a broad diffusive peak above 200 K (at high field) and a second more prominent relaxation peak at lower temperature.
The diffusive relaxation yields an activation barrier for the isolated \lip\ in the range of other measurements and calculations.
However, the range over which it is the dominant relaxation mechanism limits the ability to discern characteristic features of 1D diffusion.
In high field, the spectrum has two components, a large amplitude broad resonance and a quadrupolar multiplet that
is only revealed by an RF comb excitation\cite{ZnO2022}. The quadrupole splitting is surprisingly small and increases with temperature.
Its value provides information on the implantation site. To account for many unexpected features of the data,
it is suggested that a fraction of the implanted ion traps an electron forming the neutral \elin.

The remainder of this paper is organized as follows: \Cref{sec:experiment} describes the experiment; the results and their basic analysis is presented in 
\Cref{sec:results}. Further analysis and interpretation are presented in \Cref{sec:discussion} with a summary in \Cref{sec:summary}.
Complementary information peripheral to the main results,
such as some additional data and calculation details,
is given in \Cref{sampdep,bdep,dft,hb,trap}.

\section{Experiment \label{sec:experiment}}

The ion-implanted \bnmr\ experiments were conducted at TRIUMF's Isotope Separator and Accelerator facility which provides
low energy ($1-30$ keV) ion beams of short-lived radioisotopes for materials research.
Here we use \elip\ with nuclear spin $I = 2$, a half-life of $\tau_{1/2} = 848$ ms, a gyromagnetic ratio $\gamma = 2\pi \times 6.3019 $ MHz T$^{-1}$,
and quadrupole moment of $+$32.6 mb \cite{2011-Voss,2015-MacFarlane-SSNMR-68-1}.
The beam is transported through a high vacuum beamline and polarized in-flight using a fast collinear optical pumping scheme
using circularly polarized light. The direction of polarization (helicity)
is selected by the handedness of the light polarization. In the measurement, helicity is alternated systematically with data
collected and stored separately. 
At the end of the beamline, the \elip\ is directed into one of two NMR spectrometers where it is implanted into the sample in a differentially pumped vacuum space with a typical background pressure of $10^{-10}$
Torr. The spectrometer provides a static magnetic field $B_0$ along the spin polarization, as well as a transverse radiofrequency (RF) field.
The NMR signal is detected via the parity violating weak $\beta$-decay which correlates the emission direction of the high energy $\beta$ electron
with the nuclear spin at the instant of decay. This is accomplished with two fast plastic scintillators (that detect
the emitted $\beta$s with a high efficiency) located on opposite sides of the sample along the polarization direction $z$.
For example, in the high field spectrometer, a static field in the range of several T is applied along the beam direction by a high homogeneity superconducting solenoid.
The scintillation detectors are the ``forward" ($F$) counter, downstream of the sample, and the upstream ``backward" ($B$) counter with an aperture to allow
the beam to pass through it. The beta counts $N_F$ and $N_B$ in these detectors are measured under specific conditions, and the experimental asymmetry,
defined by,
\begin{equation}
A(t) = \frac{N_{F}(t) - N_{B}(t)}{N_{F}(t)+N_{B}(t)} = \mathcal{A}_0 p_z ,
\label{eqn_asym}
\end{equation}
where $p_z$ is the polarization averaged over the ensemble of decaying spins (i.e., $p_z = \langle I_z \rangle / I$), is extracted.
The amplitude $\mathcal{A}_0$ is determined by the intrinsic asymmetry of the beta decay (1/3 for \eli), the effective detector solid angles at $B_0$, and the performance of the optical polarizer.

Three types of measurements were performed: 1) \slr; 2) resonance, and 3) a quadrupolar resonance comb. In 1), the \elip\ beam is pulsed (4 s pulse)
using an electrostatic kicker, followed by a (12 s) beam-off period. This cycle repeats, and the time-resolved $\beta$-decay counts are monitored with a time-resolution of 10 ms.
Data from subsequent pulses are combined to obtain higher statistics, for a typical total run time of $\sim$30 min.
In 2), the \elip\ is continuously implanted while the single tone \rf\ frequency is stepped slowly through a range of frequencies around the NMR frequency
$\omega_0 = 2\pi \nu_0 = \gamma B_0$ using a frequency step that is a small fraction of the linewidth.
At each step, the $\beta$ counts are accumulated over a short integration period (1 s).
The scan is then repeated, alternating both the polarization direction as well as the frequency sweep direction, to minimize systematics and increase statistics.
The scans are then combined to yield the measured spectrum. When the RF frequency matches $\nu_0$, it causes rapid spin precession about the RF field,
and the asymmetry is reduced.
In 3), the \rf\ is an equal amplitude comb, the sum of four equal amplitude sine waves at frequencies: $\tilde{\nu}_0 \pm \tilde{\nu}_q$ and $\tilde{\nu}_0 \pm 3\tilde{\nu}_q$.
The fixed parameter $\tilde{\nu}_0$ is chosen as the center of the resonance from 2),
and the comb splitting parameter $\tilde{\nu}_q$ is scanned over a defined range (alternating helicity and frequency step direction).
For a resonance split into the $2I = 4$ quadrupolar satellites,
the comb can simultaneously excite all the single quantum transitions when $\tilde{\nu}_q$ matches the quadrupole frequency $\nu_q$,
strongly enhancing (up to $\sim10\times$) the signal amplitude\cite{ZnO2022}.
This is particularly useful for well-defined quadrupolar lines from a specific crystallographic site that would be difficult
or impossible to detect in the presence of other overlapping lines.

\begin{figure}%[tbh]
\centering
\includegraphics[width=\columnwidth]{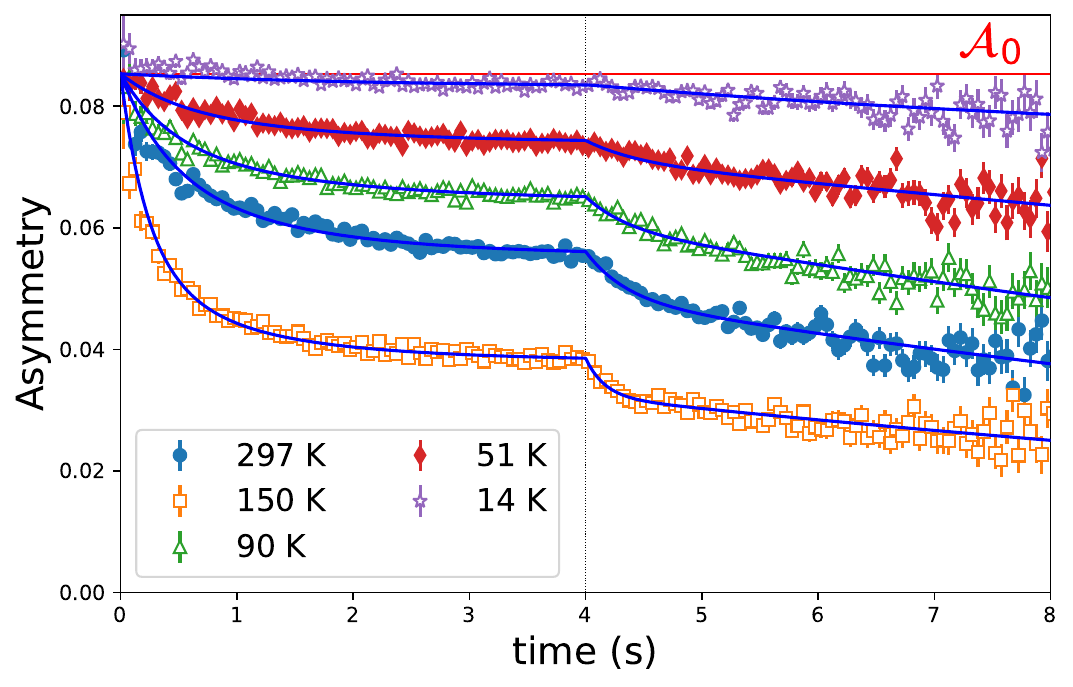}
\caption{The experimental beta decay asymmetry as a function of time showing the temperature dependence of the spin-lattice relaxation of \elip\ implanted into quartz with $B_0 = 6.55$ T $\parallel$ $c$.
The vertical line marks the end of the beam pulse. The relaxation is fastest at an intermediate temperature and shows a biexponential character (blue curves) with
a fast relaxing fraction that evolves with temperature. The data are shown with 50 ms bins for clarity.
Note the time dependence of the statistical error bars reflects the radioactive lifetime of \eli.  
}
\label{fig:biexp-ex-6p55}
\end{figure}

The data is taken on a hydrothermally grown commercial single crystal of $\alpha$-SiO$_2$ (SurfaceNet, Rheine, Germany).
The colorless transparent crystal measuring $8\times10\times0.5$ mm was oriented with its large flat
face perpendicular to the $c$-axis which was ``epi-polished''
with a surface roughness $\sim 1$ nm. The Al content is on the order of 5 ppm or less,
and the main impurity is hydrogen which can be as high as 0.25\% relative to Si\cite{Brice1985}.
In \Cref{sampdep}, we also include a comparison with an ``N grade'' fused quartz sample (Tosoh, Tokyo, Japan) at room temperature.
In the course of these measurements, we also observed {\it scintillation} in quartz at low temperature
consistent with its known ionoluminescence\cite{Bettiol1997}.

The primary data at $B_0=6.55$ T $\parallel c$ uses an implantation energy of 25 keV, but we also include data from lower fields up to 22 mT $\perp c$ at 28 keV.
SRIM\cite{Ziegler2010} simulations predict a mean range of 163(181) nm with a straggle of 55(59) nm at 25(28) keV.
However, the implantation direction is along $c$ in both cases, and in view of the structure shown in \Cref{fig:channel},
there is likely a channelled fraction that penetrates considerably further than the simulations predict.
We also note that, compared to typical stable isotope ion beams, the radioactive \elip\ beam is very low intensity.
The maximum beam rate was $\sim 5 \times 10^6$ ions per second in a beamspot about 2 mm in diameter
(for a flux $\sim 2 \times 10^8$ ions/cm$^2$/s).
The total fluence over the course of the experiment was $\sim 5.5 \times 10^{12}$ ions/cm$^2$.

\section{Results\label{sec:results}}

\subsection{Spin-Lattice Relaxation\label{subsec:SLR}}

Examples of the high magnetic field relaxation data are shown in \Cref{fig:biexp-ex-6p55}, exhibiting the bipartite relaxation
(recovery curve) due to the 4 second beam pulse.
At room temperature, the relaxation is remarkably fast for a nonmagnetic insulator.
The relaxation is not directly related to the observed scintillation. For example, 
we find an even stronger scintillation in $\alpha$-Al$_2$O$_3$, but in contrast,
the relaxation in high magnetic fields is very slow\cite{MacFarlane2023-sapphire}.
It is also strongly (and nonmonotonically) temperature dependent with the fastest relaxation in the vicinity of 150 K.
The simplest relaxation function that can fit the data is a biexponential where the spin polarization at time $t$ after implantation follows,
\begin{equation}
p_z(t) = (1-f_{f})e^{-\lambda_{s}t}+f_{f}e^{-\lambda_{f}t},
\label{eqn_bi_exp}
\end{equation}
with about a factor of 10 between the two \slr\ rates $\lambda_i \equiv (1/T_1)_i$.
The data were fit to this function convoluted with the square beam pulse with a common $T$-independent
initial asymmetry $\mathcal{A}_0 = 0.0885(1)$, consistent with the full asymmetry from calibration runs in MgO and ZnO.
In this fit, the fraction $f_{f}$ was a free parameter varying with temperature, see \Cref{fig:Fractions}a).
The resulting global fit was good (reduced $\chi^2 = 1.03$), with some examples shown in 
\Cref{fig:biexp-ex-6p55}.

\begin{figure}%[tbh]
{\centering
\includegraphics[width=\columnwidth]{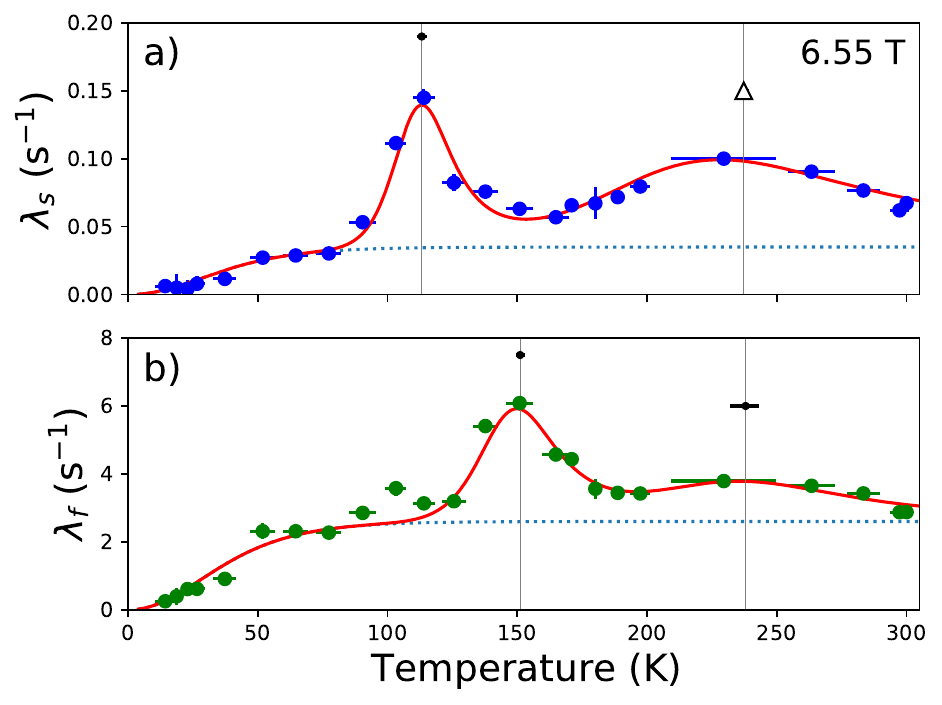}
\caption{ The a) slow and b) fast relaxation rates at 6.55 T as a function of temperature,
showing a sharper low temperature and broader high temperature peak. From parabolic fits in the vicinity of the maxima, 
the peak temperatures are shown as the black symbols and vertical lines.
The background rate (dotted line) slows below 50 K.
The red curves are a guide to the eye.
}
\label{fig:SLRrates6p55}
}
\end{figure}

The relaxation rates $\lambda_i$ from this fit are shown versus $T$ in \Cref{fig:SLRrates6p55}.
They reveal a low temperature increase to $\sim 50$ K (similar in both components)
followed by sharp peaks at different temperatures in the slow (115 K) and fast (151 K) components
and much broader peaks around 250 K in both.
To locate the peak positions, subsets of the data about each maximum were fit with a simple parabola.
The resulting best fit positions are marked by the vertical lines in \Cref{fig:SLRrates6p55},
with MINOS uncertainties shown as the horizontal errorbar in the corresponding points above the data.

\begin{figure}%[tbh]
\centering
\includegraphics[width=\columnwidth]{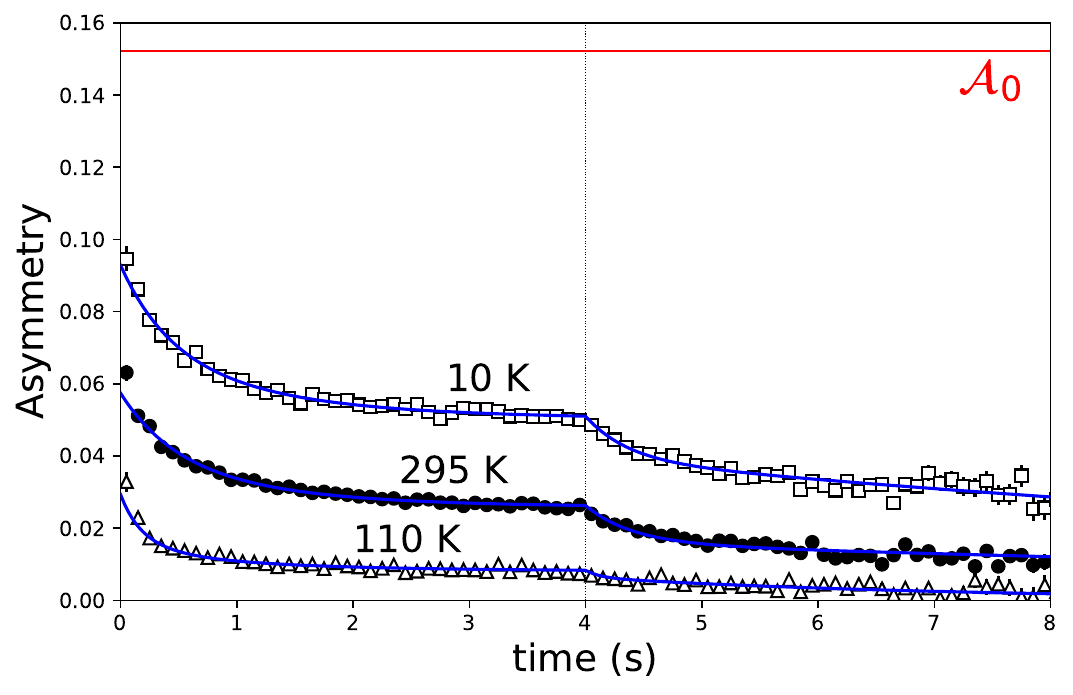}
\caption{ Spin-lattice relaxation of \eli\ at 20 mT $\perp c$.
The red line indicates the calibrated full asymmetry $\mathcal{A}_0$, clearly
demonstrating a large missing fraction at all temperatures.
The data are shown with 100 ms bins for clarity with biexponential fits.
}
\label{fig:200GSLR}
\end{figure}

In contrast, at low magnetic field there is a surprising loss of asymmetry.
The relaxation data at 20 mT as a function of temperature is shown in \Cref{fig:200GSLR}.
The expected asymmetry, determined by a calibration run, is $\mathcal{A}_0 = 0.152(2)$ shown by the red dashed line, so
there is a substantial ``missing fraction'', i.e.\ a fraction of the implanted \eli\ whose spin has relaxed so quickly that its signal is wiped-out.
The missing fraction is present at all temperatures, but it is largest in the vicinity of 100 K.

\begin{figure}%[tbh]
{\centering
\includegraphics[width=\columnwidth]{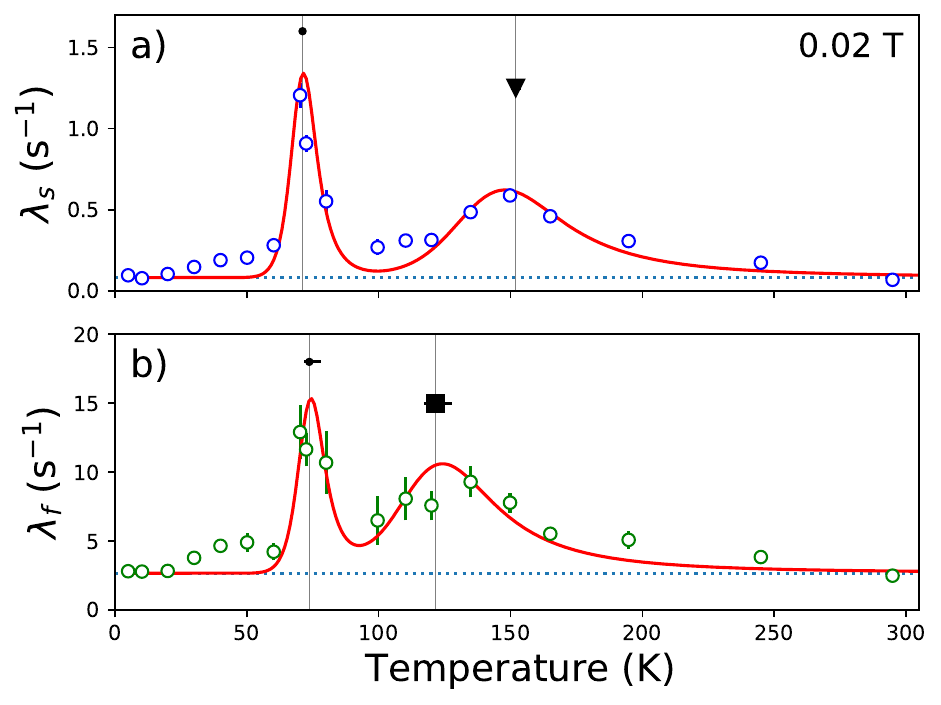}
\caption{ The a) slow and b) fast relaxation rates at 20 mT as a function of temperature exhibits features similar to
high field. From parabolic fits, 
the peak temperatures are shown as the black symbols and vertical lines.
The constant background rate is shown as the dotted line, and the red curves are a guide to the eye.
}
\label{fig:SLRrates02}
}
\end{figure}

Based on the high field data and the field dependence of the relaxation at room temperature (see \Cref{bdep}),
we also fit the relaxation at 20 mT with a biexponential, but unlike the 6.55 T data, the amplitude varies with $T$.
The resulting fits have reduced $\chi^2 = 1.0$, and examples are shown as the blue curves in \Cref{fig:200GSLR}. 
While not constant, $f_s$ only varies from 29-50 \%. In contrast, the amplitude is more strongly temperature dependent
[see \Cref{sec:discussion} and the plot of the missing fraction $[\mathcal{A}_0 - A(T)]/\mathcal{A}_0$ in \Cref{fig:Fractions}b)].
From these fits, the $\lambda_i(T)$, shown in \Cref{fig:SLRrates02}, exhibit similar peaks to the high field rates.
At such a low field, other relaxation mechanisms that are quenched in high magnetic fields may
modify and even dominate the relaxation. The similarity in the $T$ dependence
suggests this is not the case. At 300 K, both rates are comparable in magnitude to 6.55 T.
However, the two components do not correspond. It is very likely that the fast relaxation at
high field becomes the missing fraction at 20 mT as suggested by the similar $T$ dependence to the fractions,
see \Cref{fig:Fractions}. Compared to $\lambda_s$ at 6.55 T, the peaks in \Cref{fig:SLRrates02}
are shifted to substantially lower temperature as would be expected for thermally activated relaxation, given the much smaller $\omega_0$.
Using a similar strategy as above, the peak temperatures (vertical lines in \Cref{fig:SLRrates02}) were determined by parabolic fits.
Aside from the peaks, the background relaxation at low field is more independent of temperature, particularly below 50 K.

In a separate measurement, we implanted 25 keV \elip\ in the temperature range $55-450$ K to test
for migration back to the surface via the online alpha radiotracer method\cite{Chatzichristos2019}.
This technique uses the pair of alpha particles also produced in the \eli\ decay chain.
The $\alpha$s are strongly attenuated by the sample, so only decays occurring very near the surface produce $\alpha$s
that are detected.
Ionic motion towards/away from the surface of even a few nanometers per $\tau_{1/2}$ produces
an observable time-dependence in the alpha particle yield.
In this case, we found no time-dependence of the alpha rates at any temperature,
suggesting that at the scale of nm, the diffusion rate of implanted \elip\ in quartz is less than 1 nm/s even at 450 K. 
However, nanoscale diffusion does not always follow the intrinsic hop rate. 
For example, if the $c$-axis channels are blocked by defects every few nanometers,
then \eli\ might exhibit very fast hopping without significant time evolution of the implantation profile.

Having summarized the relaxation data, in the next section we present the resonance spectra in high magnetic fields.

\subsection{Resonance Spectra\label{subsec:RES}}

\begin{figure}%[tbh]
\centering
\includegraphics[width=\columnwidth]{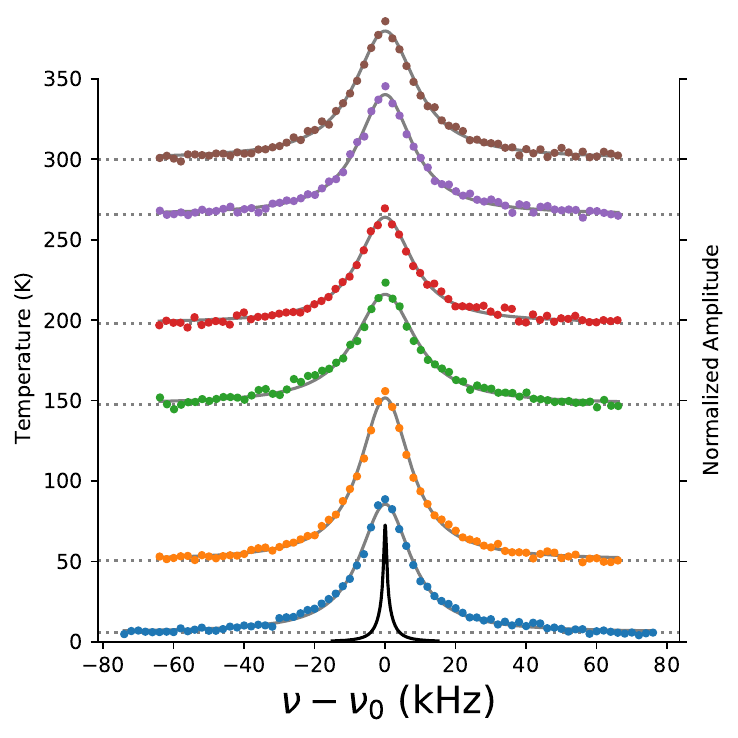}
\caption{ Single tone RF \bnmr\ spectra of \elip\ implanted in an $\alpha$-SiO$_2$ single crystal with $B_0 = 6.55$ Tesla $\parallel c$.
The large amplitude resonance is remarkably broad and not strongly temperature dependent.
The baselines are offset to match the temperature label on the left.
The black curve shows a calibration spectrum in MgO (scaled down by 2) for comparison.}
\label{fig:1f}
\end{figure}

Single tone RF spectra in high magnetic field are shown in \Cref{fig:1f}
normalized to the off-resonance time-average ``baseline'' asymmetry.
They are well-described by a single Lorentzian (fit curves) resonance near the Larmor frequency
without evident quadrupolar splitting. We estimate an \eli\ shift of about $-25(2.5)$ ppm (independent of $T$)
relative to a calibration in MgO, typical of the small \eli\ shifts in nonmagnetic oxides.
The line is unexpectedly broad, with a full-width at half-maximum (FWHM) of $\sim 20$ kHz independent of $T$.
For comparison, we include the resonance in MgO in \Cref{fig:1f} which, while it is about 10 times narrower than the quartz line, 
is still power broadened far above the instrumental limit\cite{Fujimoto2019}. 
The baseline asymmetry (and hence the resonance amplitude) depends on $T$ through the SLR rate, so without normalization,
the resonances are smaller in the region of the SLR peaks.
At all $T$, the resonance amplitude is $\sim 40$ \% of the baseline asymmetry (full saturation).
This is remarkably large for the breadth of the line and strongly suggests that the resonance is not a static lineshape.
At room temperature, this is confirmed by a time-resolved hole-burning measurement that indicates a substantial
effect of slow spectral dynamics of the \eli, i.e.\ the resonance condition is fluctuating on the timescale of 10 ms.
This enhances the effect of the RF, amplifying the time-integrated resonances in \Cref{fig:1f}, as well as modifying the line shape.

\begin{figure}%[tbh]
\centering
\includegraphics[width=\columnwidth]{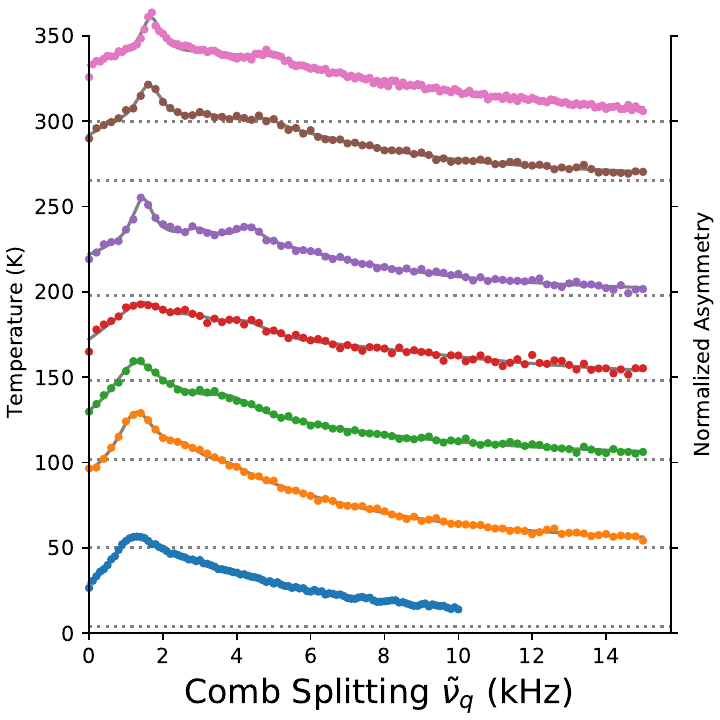}
\caption{ Quadrupolar RF comb \bnmr\ spectra of \elip\ implanted in an $\alpha$-SiO$_2$ single crystal at $B_0 = 6.55 \parallel c$ Tesla.
The comb is centered on the resonance in \Cref{fig:1f} and the ordinate is the comb splitting parameter $\tilde{\nu}_q$.
When this matches the characteristic splitting of an \eli\ site, a substantially enhanced resonance is observed.
The baselines are offset to match the temperature label on the left.}
\label{fig:1w}
\end{figure}

\begin{figure}%[tbh]
\centering
\includegraphics[width=\columnwidth]{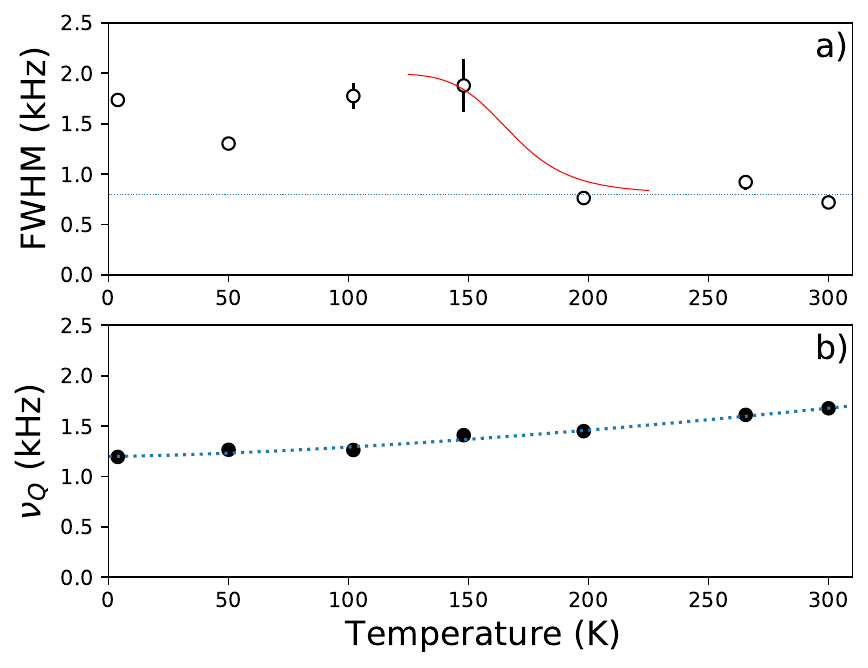}
\caption{ Temperature dependence of a) the linewidth of the narrow comb resonance and b) the position of the comb resonance $\nu_q$ from fits to the spectra shown in \Cref{fig:1w}.
The red curve shows a possible motional narrowing that would account for the drop in width above 150 K.
The dashed curve in b) is the fit described in the text.}
\label{fig:1w-fits}
\end{figure}

Though no quadrupole splitting is evident in \Cref{fig:1f}, with the four frequency comb excitation,
we find a resonance at finite splitting $\tilde{\nu}_q$.
With the comb centred on the Lorentzian in \Cref{fig:1f}, spectra as a function of comb splitting parameter $\tilde{\nu}_q$ 
are shown in \Cref{fig:1w}. As the comb lines sweep through the broad spectrum in \Cref{fig:1f}, there will be a similarly broad signal
centred at $\tilde{\nu}_q=0$, and this is the case in \Cref{fig:1w}. However, superposed on this broad line, there
is a resonant signal near 2 kHz together with its alias at three times this frequency.
The four line satellite pattern corresponding to this value of $\nu_q$ overlaps the centre of the broad resonance in \Cref{fig:1f}
and, without the comb enhancement, is too weak to detect using the single tone RF.
The comb resonance in \Cref{fig:1w} is broad at low temperature but narrows substantially above 150 K where the \elip\ should be mobile.
Careful inspection of the single tone spectra in \Cref{fig:1f} suggests a small
very sharp component superposed on the broad Lorentzian in this high temperature range.
This may be the multiquantum resonance at the centre of the narrowed quadrupolar multiplet giving the comb resonance.
Fitting the comb spectra to a sum of Lorentzians: narrow lines for the primary resonance and its aliases (see discussion) and a broad background
(FWHM $\sim 10$ kHz) corresponding to the broad dynamic Lorentzian line in \Cref{fig:1f} (curves in \Cref{fig:1w}),
we obtain the line width and quadrupole frequency $\nu_q$ plotted as a function of $T$ in \Cref{fig:1w-fits}.

\section{Discussion \label{sec:discussion}}

\begin{figure}%[tbh]
\centering
\includegraphics[width=\columnwidth]{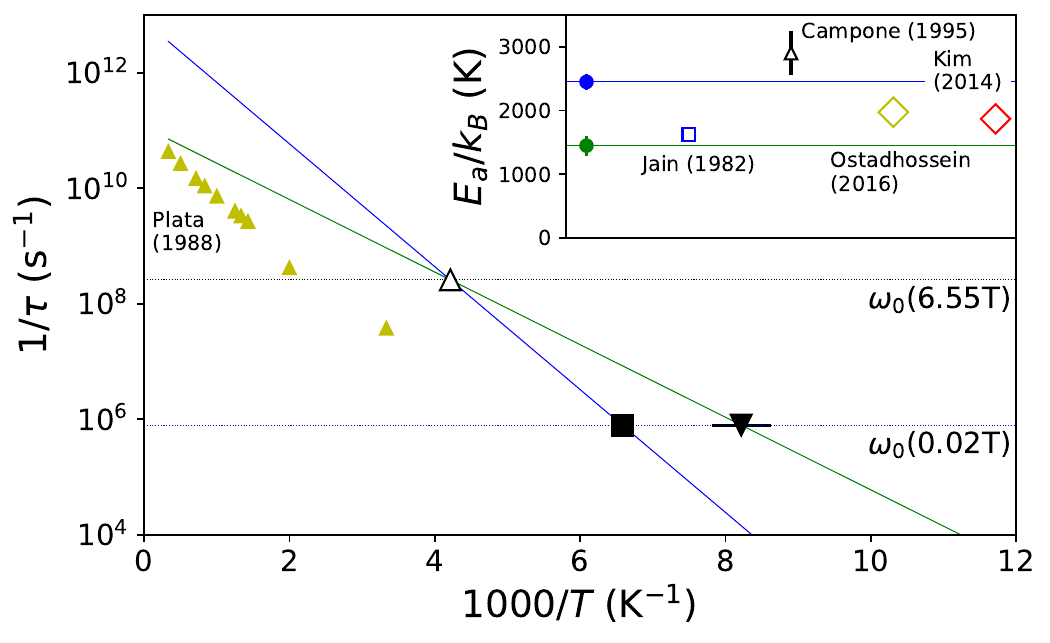}
\caption{\elip\ hop rates determined from the high temperature SLR peak positions at the two fields
(giving the NMR frequencies shown as horizontal lines) [open triangle from \Cref{fig:SLRrates6p55}a), closed symbols
from \Cref{fig:SLRrates02}] with the resulting Arrhenius dependencies shown as the blue and green lines.
Yellow triangles are calculated rates from Ref.\ \onlinecite{Plata1988}. 
{\bf Inset:} The barrier to hopping for \lip\ in \quartz\ from the slopes in the main panel compared to values
from the literature. The large diamonds are calculated\cite{Kim2014,Ostadhossein2016},
while others are experimental\cite{Jain1982,Campone1995}.
}
\label{fig:Arrhen}
\end{figure}

We begin with the high temperature regime, where we expect the implanted \elip\ to be diffusing.
Accepting the conclusion that \lip\ motion is frozen below 150 K\cite{Weil1984}, we attribute the broad high temperature peaks in the
SLR rates in \Cref{fig:SLRrates6p55,fig:SLRrates02} to hopping of \elip\ along the channel.
Assuming a simple BPP picture, the peak position occurs when the hop rate matches the Larmor frequency.
At 6.55 T, the peaks in both fast and slow fractions are practically coincident. 
Presuming the fast component is missing at 20 mT, both fast and slow
components at 20 mT in \Cref{fig:SLRrates02} correspond to the high field slow component.
The peak positions (temperatures) from the parabolic fits are shown on an Arrhenius scale in \Cref{fig:Arrhen}.
Treating the two components at 20 mT as independent measurements yields two different Arrhenius laws
with barriers plotted in the inset of \Cref{fig:Arrhen}.
The prefactors (vertical intercept in \Cref{fig:Arrhen}) are in the expected range
[$8(5)\times 10^{12}$ (slow) and $1.2(8)\times 10^{11}$ (fast)],
and thus do not exhibit any significant anomaly\cite{Villa1983}.
The slopes ($-E_a/k_B$) are in the range found in other measurements\cite{Jain1982,Campone1995}
and calculations\cite{Kim2014,Ostadhossein2016} for the isolated \lip.
It is difficult to obtain an absolute value of $1/\tau$ from the conductivity due to uncertainty in the ionic carrier concentration.
Similarly, the prefactor $1/\tau_0$ is difficult to calculate, but Ref.\ \onlinecite{Plata1988} does report absolute
theoretical hop rates at higher temperatures (yellow triangles in \Cref{fig:Arrhen}). From these, one can see the slope
is comparable to the steeper of our measurements, but the prefactor is an order of magnitude smaller. 
Though the two component character of the low field relaxation is clear in \Cref{fig:200GSLR},
we do not know what distinguishes the components. Combined with substantial uncertainty in locating the relaxation peaks
in \Cref{fig:SLRrates02}, we report the average of the
two slopes
[178(43) meV or \qty{2.1 \pm 0.5 e3}{\kelvin}]
in \Cref{fig:Arrhen} as our best estimate of the hopping barrier for \elip\ in quartz.  

Quite generally, the temperature and field dependence of the diffusive SLR rate is given by\cite{Heitjans2012,Beckmann1988} 
\begin{equation}
\lambda_{\mathrm{diff}} = C \left[ J(\omega_0,\tau)  + 4 J(2\omega_0,\tau) \right],
\label{eq:slrr}
\end{equation}
where $C$ is a scale factor determined by the nuclear spin coupling, and $J$ is the fluctuation spectral density.
The temperature dependence of $\lambda(T)$ arises from the correlation time $\tau(T)$ which we assume to follow an 
Arrhenius dependence as in \Cref{eq:arr}.
In the case of an exponentially decaying autocorrelation, we have the BPP spectral density\cite{BPP1948} appropriate to 
3D diffusion,
\begin{equation}
J_{\mathrm{BPP}}(\omega,\tau) = \frac{2 \tau}{1 + (\omega \tau)^2}.
\label{eq:bppspecdens}
\end{equation}
This does not have the correct asymptotic behavior for 1D diffusion as $\omega \tau \rightarrow 0$\cite{Sholl1981,Richards1979}.
However, this is remedied by a simple heuristic factor,  
\begin{equation}
J_{\mathrm{1D}}(\omega,\tau) = \sqrt{1 + \frac{1}{\omega \tau}} \cdot J_{\mathrm{BPP}}(\omega,\tau) .
\label{eq:specdens1D}
\end{equation}
Both spectral densities produce a relaxation peak when $\omega \tau(T) \sim 1$. Beyond this, the width in temperature is determined by how fast the Arrhenius $1/\tau$ sweeps through the
peak region. A larger prefactor $1/\tau_0$ requires a larger $E_a$ and results in a narrower peak, while
1D diffusion yields a characteristically asymmetric peak with a weaker fall off at high temperature\cite{Heitjans2012},
a feature captured by \Cref{eq:specdens1D}.

\begin{figure}%[tbh]
\centering
\includegraphics[width=\columnwidth]{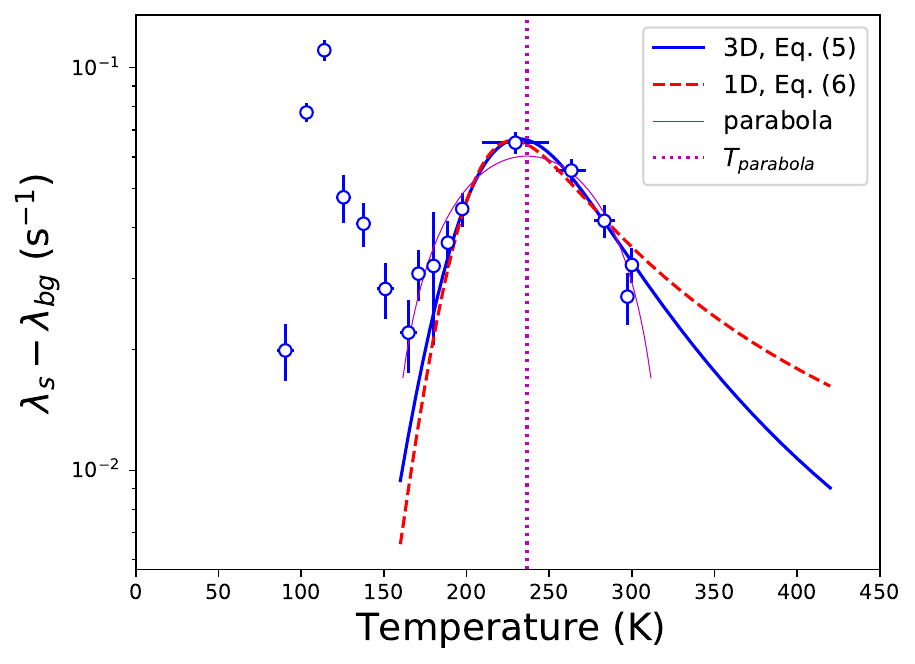}
\caption{ The relaxation rate of the slow component at $B_0=6.55$ T with the background rate subtracted compared to calculated model
diffusive rates from \Cref{eq:slrr} (curves) using the BPP spectral density \Cref{eq:bppspecdens} (blue) and the heuristic 1D function \Cref{eq:specdens1D} (red dashed).
Also shown is the parabolic fit (purple) centred on the vertical dotted line which is used in \Cref{fig:Arrhen}.
}
\label{fig:BPP-Comp}
\end{figure}

To compare these models to the data, we select the slow component at high magnetic field,
which has the largest variation and the two peaks in
$\lambda_s(T)$ are best separated. Subtracting the background rate (constant above 100 K),
we get the relaxation rates shown in \Cref{fig:BPP-Comp}. The two models each depend on 3 parameters:
$C$ in \Cref{eq:slrr} and the Arrhenius parameters. The latter two are highly correlated,
so while the curves are not fits, they indicate that both models \emph{could} fit the data quite well.
The corresponding barriers are low [$E_a/k_{B} = 1430$ (3D) and 1650 K (1D)] and
the intercepts required to match the peak width are $1/\tau_0 = 2$ and 9 $\times 10^{11}$ s$^{-1}$,
somewhat less than the analysis using both fields shown in \Cref{fig:Arrhen}.
From \Cref{fig:BPP-Comp}, it is clear that distinguishing 1D diffusive relaxation
requires a greater dynamic range of $\lambda$ and a wider $T$ range.
Though the detailed shape of the peak (e.g., its width and symmetry) contain more information than the peak position alone,
the ability to discern features characteristic of 1D is limited here by the other relaxation channels that
give rise to the low temperature peak and the background rate. Given these limitations, the peak position analysis presented above
remains a reasonable model-independent approach. With more beam time, the Arrhenius parameters could be 
better determined by measurements at several more magnetic fields.

A significant conclusion that can be drawn directly from the SLR data is that diffusive relaxation is not the only operative mechanism.
From \Cref{fig:SLRrates6p55,fig:SLRrates02}, it is clear that both fast and slow rates have similar $T$ dependencies,
suggesting they share a similar origin, with the distinction between $\lambda_f$ and $\lambda_s$ mostly one of scale.
It appears reasonable to decompose the relaxation into three distinct processes,
\begin{equation}
\lambda(T) = \lambda_{\mathrm{diff}} + \lambda_{\mathrm{pk}} + \lambda_{\mathrm{bg}}, 
\label{eq:slr-decomp}
\end{equation}
where $\lambda_{\mathrm{diff}}$ accounts for the high $T$ broad diffusive peak discussed above
[\Cref{eq:slrr}], $\lambda_{\mathrm{pk}}$ is the sharp lower $T$ peak, and $\lambda_{\mathrm{bg}}$ is a background rate that
is independent of $T$ at low field, while at high field, it increases from a small low $T$ value to
plateau above $\sim 50$ K (dotted lines in \Cref{fig:SLRrates6p55,fig:SLRrates02}).
The background rate is reminiscent of NMR relaxation due to
dilute paramagnetic centers in wide band gap insulators,
where it is due to the fluctuations of the unpaired electronic spin of the defects\cite{Persyn1965,Thompson1969}.
At low $T$,  the fluctuations are so slow that they are ineffective at
relaxing the nuclei in high $B_0$, but the fluctuation rate $1/T_{1e}$ increases via spin-lattice coupling
with the thermally excited phonons until the low frequency tail of its spectral density overlaps the NMR frequency
and it begins to cause relaxation. At low field, this can occur at a much lower $T$. The high $T$ plateau is not as well understood,
but it is likely a superposition of several broad features\cite{Persyn1965,Thompson1969}.
We will return to consider the origin of the remaining term $\lambda_{\mathrm{pk}}$ below.

To obtain a more detailed understanding of the diffusive dynamics and to assist in interpreting the unexpected features of the data at lower $T$, 
we next consider the site of the implanted \elip.
In a crystal, the implanted ion generally stops in a vacant interstitial site, often a high symmetry crystallographic site.
In some cases, multiple inequivalent sites are populated, some being metastable with a barrier to the lowest energy site in the unit cell.
The NMR spectrum contains information about the site, particularly through the quadrupolar interaction which couples
the $(I>1/2)$ spin of the nucleus to the local electric field gradient (EFG) tensor.
The EFG is determined by the charge distribution surrounding the site, and the scale of the coupling is
given by the product of the nuclear electric quadrupole moment ($eQ$) with the principal component of the EFG ($V_{zz} = eq$) defining
the quadrupole frequency\cite{CohenReif1957} for \eli\ as,
\begin{equation}
\nu_q = \frac{e^2qQ}{4h}.
\label{eq:nuq}
\end{equation}
Usually this is much smaller than the Zeeman interaction with the applied field $B_{0}$
(i.e., $\nu_q \ll \nu_0$),
and the
quadrupolar coupling can be treated as a perturbation, which to first-order, splits the NMR line into a symmetric multiplet of $2I$ quadrupolar satellites
corresponding to the $|\Delta m| = 1$ transitions among the $2I+1$ states of the nuclear spin. For the present case, 
the satellite frequencies are given by
\begin{equation}
\nu_i = \nu_0 + n_i \frac{\nu_q}{4} \left( 3 \cos^2 \theta - 1 + \eta \sin^2 \theta \cos 2\phi \right),
\label{eq:splittings}
\end{equation}
where $n_i = \pm 1 (\pm 3)$ for the inner(outer) lines; $\eta \in [0,1]$ is the dimensionless asymmetry parameter of the
EFG; and $\theta$ and $\phi$ are the polar and azimuthal angles of $B_0$ in the principal axis coordinate system of the EFG.
This interaction is often the largest coupling of a quadrupolar nuclear spin to its crystalline environment.
The RF comb spectra show resonances at finite splitting $\tilde{\nu}_q$
(see \Cref{fig:1w}),
yielding the $\nu_q$ characteristic of the corresponding \eli\ site.
The measured value is much smaller than in either Al$_2$O$_3$\cite{MacFarlane2023-sapphire} or ZnO\cite{ZnO2022}.
In part, this is due to the higher covalence of SiO$_2$ that makes the charge distribution more uniform.
The $T$ dependence of $\nu_q$ [plotted in \Cref{fig:1w-fits} b)] shows a clear {\it increase} with temperature,
in contrast to the more common case of a decrease. The fit shown is  $\nu_q(T) = \nu_{q0} (1+ BT^{1.5})$ with the zero temperature quadrupole frequency
$\nu_{q0} = 1.199(12)$ kHz and $B=7.7(3) \times 10^{-5}$ K$^{-1.5}$.
An increasing EFG is not unprecedented\cite{Nikolaev2020}, but the relative change is quite large.
This trend bridges the transition to mobility marked by the line narrowing, so it does not appear to be directly related to motional averaging.
The $T$ dependence of $\nu_q$ is primarily due to lattice vibrations, and the unusual variation found here may arise from the
flexibility of the quartz structure to torsion of the SiO$_4$ tetrahedra\cite{Sartbaeva2005}.

\begin{figure}%[tbh]
\centering
\includegraphics[width=\columnwidth]{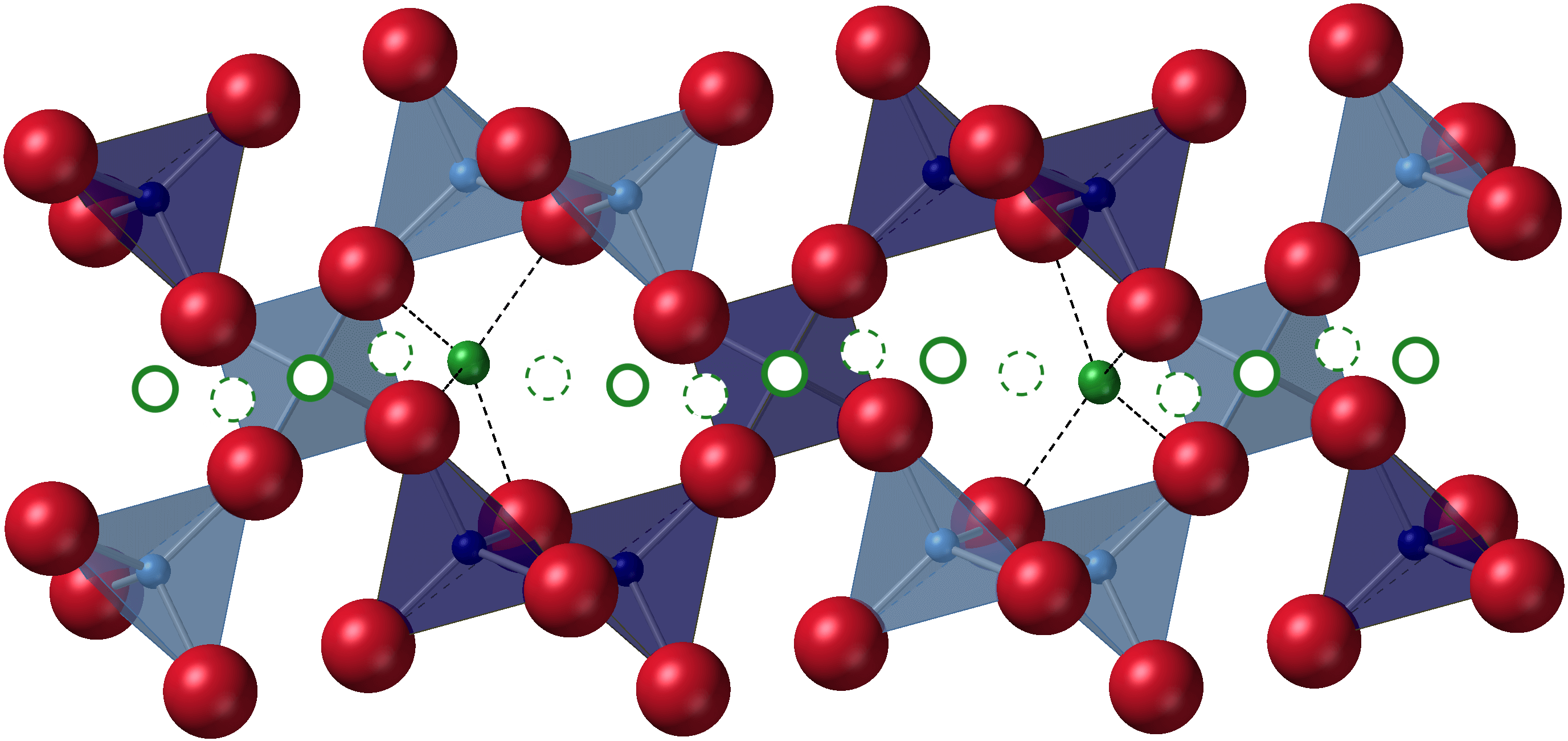}
\caption{ Side view of the $c$-axis channel in quartz showing occupied interstitial Li sites (green spheres),
vacant sites (solid green circles). The dashed green circles indicate the metastable Li sites midway between adjacent stable sites.
There are three minimum energy Li sites in each channel per $c$-axis lattice constant
on the $2$-fold axes $\perp c$ of the ideal structure.
This axis rotates by $120^\circ$ about $c$ between adjacent sites.
The double spiral of the linked SiO$_4$ are shown as the light and dark blue tetrahedra.}
\label{fig:barrier}
\end{figure}

Naturally occurring \lip\ impurities in $\alpha$-SiO$_2$ occupy interstitial sites in the $c$-axis channels
(see \Cref{fig:channel}).
EPR of trapped holes\cite{Weil1984} and electrons\cite{Jani1986} in the vicinity of interstitial \lip\ show hyperfine
splittings from the $^7$Li nuclear spin.
The site determined by analyzing these splittings\cite{Weil1984,Wilson1986} is near the central axis of the channel
[$c$-axis edge of the unit cell in \Cref{fig:channel}]
where it intersects the two-fold crystallographic axis ($\perp c$) that bridges the channel joining two Si atoms on opposite sides
(abbreviated below as TFA). This site is quasitetrahedrally coordinated by oxygen bonded to the two Si\cite{Weil1984}.
There are three such sites in the unit cell, equally spaced along $c$,
but with the bridging TFA rotated by $120^\circ$ about $c$. The array of such sites is not linear
but describes a tight helix winding along $c$\cite{Plata1988},
as illustrated by the filled green circles and spheres in \Cref{fig:barrier}.
Without the charge of the unpaired electron (that gives the EPR signal), the site of {\it isolated} \lip\
is difficult to determine experimentally\cite{Wahl1997} and may
differ, but calculations find a similar quasi-tetrahedral in-channel site\cite{Cora1994,Kim2014,Ostadhossein2016,Sartbaeva2004}.
The EFG at the nucleus is predominantly due to the nearby charge distribution, as it falls as $r^{-3}$.
Moreover, it is zero by symmetry when the charge distribution is cubic.
Thus, qualitatively, one can expect the quasi-tetrahedral site to have a small EFG.

Using supercell density functional theory (DFT) calculations [see \Cref{dft} for details], we confirm the lowest energy site is as described above.
However, midway between neighboring minimum energy sites, there is another shallow minimum $\sim 0.13$ eV higher
in energy which is also quasi-tetrahedrally coordinated by oxygen, suggesting the hopping barrier may include a metastable transition state.
This may be the `B’ transition site referred to in Ref.\ \onlinecite{Kim2014}.
The stable site is a Wyckoff $3a$ site along the TFA,
with a $\sim 0.12$ \AA\ transverse offset from the channel center. 
Along the channel, the metastable site is midway between adjacent stable sites and has a similar transverse offset $\sim 0.11$ \AA,
defining the radius of the helical path depicted in \Cref{fig:barrier}.

\begin{table}%[tbh]
\caption{ {\sc gipaw} calculated quadrupole parameters for the dilute \elip\ from the $3\times3\times4$ supercell in the two sites.}
\centering
\begin{ruledtabular}
\begin{tabular}{c c c}
Site & $\nu_q$ (kHz) & $\eta$\\ %\hline \hline
%\rule{0pt}{3ex}
\hline
\rule{0pt}{2.6ex}
Stable & $-29$ &  $0.50$          \\
Metastable & $-58$ &  $0.12$     \\
%\hline
\end{tabular}
\end{ruledtabular}
\label{tab:dftq}
\end{table}

From the relaxed structure at these two sites, we use {\sc gipaw}\cite{Varini2013,gipaw2001} to estimate the EFG at the Li nucleus.
This method reproduces the EFG at $^{17}$O in SiO$_2$ quite well\cite{Profeta2002,Spearing1992}.
The calculated quadrupole parameters for dilute \elip\ are summarized in \Cref{tab:dftq}.
While the EFG at the Li site is $\sim 50\times$ smaller than at oxygen, $\nu_q$ is still much larger than
than the experimental value. Comparison of {\sc gipaw} EFGs with experiment over a broad range of materials show
substantial scatter, particularly for light elements, and there is an effective uncertainty on the calculated values in 
\Cref{tab:dftq} on the order of 50 kHz\cite{Reina2025}.
Based on this, the observed comb resonance is consistent with the stable channel site.
The site symmetry determines the EFG orientation which is reliably calculated by {\sc gipaw}.
For the stable site, the EFG principal axis is along the TFA rotating by 120$^\circ$ about $c$
from one site to the next along the channel.
This means that, for the high field spectra shown in \Cref{eq:splittings},
$\theta = \qty{90}{\degree}$.
With a very small barrier to stability, the metastable site probably has a negligible population on average.
However, it may be occupied briefly by the hopping ion.
Here the EFG is about twice as large (with a smaller $\eta$), and it is no longer in the $ab$ plane.
It would be interesting to revisit the calculations of Ref.\ \onlinecite{Plata1988} including the metastable position to see if better
agreement with the experiment could be obtained.

\begin{table}%[tbh]
\caption{ Spin-bearing nuclei in the $\alpha$-SiO$_2$ crystal.
For each isotope, its concentration (relative to Si), the magnitude of its nuclear dipole moment $\mu$ (in nuclear magnetons $\mu_{N}$), along with its spin quantum number $I$, is listed.
Note that the quoted \eli\ concentration corresponds to its estimated maximum value during the experiment.}
\centering
\begin{ruledtabular}
\begin{tabular}{c c c c} 
Isotope & Concentration (ppm) & $\mu$ ($\mu_N$) & Spin $I$\\ %\hline \hline
%\rule{0pt}{3ex} 
\hline
\rule{0pt}{2.6ex} 
$^{29}$Si & $\sim 47000$  &       $-0.555$ & 1/2             \\
$^{17}$O  &   $\sim 760$  &       $-1.894$ & 5/2            \\
$^{1}$H   &  $ < 2500$   &       $+2.793$ & 1/2           \\
$^{27}$Al &   $< 5$        &       $+3.642$ & 5/2             \\
$^{8}$Li &   $< 0.0001$    &       $+1.653$ &  2             \\
%\hline
\end{tabular}
\end{ruledtabular}
\label{tab:nuclei}
\end{table}

The best resolved comb spectra in \Cref{fig:1w} are at high $T$.
The low $T$ spectrum, where \eli\ is frozen, is still peaked at $\nu_q$ but is substantially broader.
In part, this is due to the 3 distinct orientations of the EFG in the 3 randomly populated stable sites,
resulting in 3 distinct EFG angles in \Cref{eq:splittings} which split the resonances further.
This is analogous to the 3-fold splitting of the $^{29}$Si NMR from 3 distinct orientations of the SiO$_4$ tetrahedra\cite{Spearing1992}.
Another unrelated source of broadening originates in the magnetic interaction with other spin-bearing nuclei.
This ``dipolar coupling'' also broadens the resonance to an extent determined
by the abundance and magnitude of the nuclear magnetic moments.
In SiO$_2$ these are very dilute (see \Cref{tab:nuclei}), so dipolar broadening is not significant.
For example, at room temperature, the $^{29}$Si NMR linewidth is only $\sim 0.35$ kHz\cite{Spearing1992}, notably much less than our widths.

Upon warming,
the comb resonance becomes narrower and better defined above 150 K
(see \Cref{fig:1w-fits}),
suggesting the corresponding fraction of \eli\ is mobile.
In the fast diffusion limit, the EFG is averaged and there is a single quadrupole multiplet. Motional narrowing occurs 
when the rate of site-to-site hopping exceeds the frequency scale of the broadening (kHz).
At even higher $T$, when hopping is much faster, it will cause the SLR rate peak $\lambda_{\mathrm{diff}}$.
The motional SLR is almost certainly quadrupolar (from the random modulation of the quadrupole interaction due to hopping).
Such relaxation is generally anisotropic, since the relevant fluctuating terms depend on the direction of $B_0$,
so they should be different in the 20 mT data.
The square magnitude of the fluctuating coupling determines $C$ in \Cref{eq:slrr} and thus the height of the SLR peak.
Indeed, the observed heights require a fluctuating coupling on the kHz scale,
consistent with the observed $\nu_q$.

In contrast, the broad resonance in \Cref{fig:1f} shows no substantial change over a temperature range that spans this onset of mobility,
so this fraction is not diffusing. Its amplitude is enhanced dynamically by spectral diffusion on the second timescale [see \Cref{hb}].
The linewidth is nearly the same at 5 K (lowest spectrum in \Cref{fig:1f}) and room temperature.
The lack of $T$ dependence to the resonance suggests the spectral dynamics are not thermal. Rather, they are probably 
caused by implantation-related excitations. Among these, STE are quite long-lived\cite{Trukhin1992},
but the RIC persists much longer\cite{Hughes1975, Jain1982}. Fluctuations from more distant long-lived excitations
might account for the slow fluctuations seen by this immobile fraction of the implanted \elip.

\begin{figure}%[tbh]
\centering
\includegraphics[width=\columnwidth]{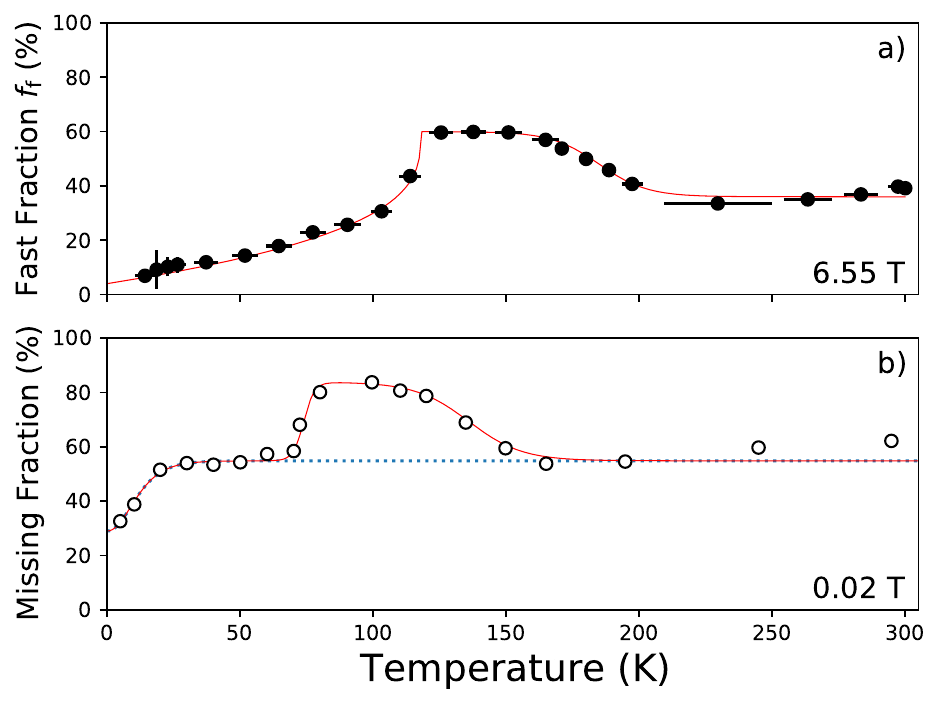}
\caption{a) The fraction of the signal that relaxes rapidly, $f_f$ at 6.55 T as
a function of temperature from the global biexponential fit.
b) The missing fraction at 0.02 T as a function of temperature. 
The red curves are empirical fits.}
\label{fig:Fractions}
\end{figure}  

Both the relaxation and resonances are multicomponent. The high field biexponential suggests two distinct populations of
\eli\footnote{The biexponential is unrelated to the common circumstance in solid state NMR where only one quadrupole satellite can be irradiated by an RF pulse.
Here, the SLR measurements use no RF, and the initial spin state has very pure dipolar polarization from the optical pumping.
Nor is it due to slow quadrupolar fluctuations\cite{Becker1982}, which give rise at most to a very small fast component (of a few \%).},
but the low field missing fraction shows that there are at least three.
The sharp comb resonance in \Cref{fig:1w} coexists with the large broad resonance in \Cref{fig:1f}.
The latter is substantially wider than the entire multiplet pattern at the comb $\nu_q$, so it must correspond to a distinct environment.
Hole-burning shows the broad resonance in \Cref{fig:1f} corresponds to the slow relaxing component; however, 
since it does not completely depolarize the remaining signal, there are also three components in the resonances: the comb resonance,
the broad dynamic resonance and another (slow-relaxing) component unaffected by the resonant RF.
Both the rates [\Cref{fig:SLRrates6p55}] and the relative fractions [\Cref{fig:Fractions} a)] evolve with $T$.
The temperature dependence of the fraction $f_f$ is considerably more complicated than for
a thermally activated site change\cite{TwoSite2009}.
More complex $T$ dependence can arise when there are multiple distinct final state defect configurations.
For example, when the positive muon $\mu^+$ is implanted into semiconductors it can form muonium
---
an isolated atomic defect (H atom analog)
---
that can exhibit distinct crystallographic and charge states that evolve with $T$\cite{Hitti1999,Baker2014}.
Another well-known phenomenon for the muon is diffusion and trapping\cite{Hartmann1988},
where dilute static defects within the structure provide an attractive trapping potential for the implanted ion.
Our sample is sufficiently pure that the vast majority of the randomly stopped \elip\ will not be in the vicinity of defects such as an Al impurity.
Even for the much more abundant hydrogen, at the estimated maximum concentration, the average distance between H impurities along the channel is about 70 nm.
While substitutional Al acts as a trap, interstitial H$^+$ would be a channel blocker. There are a number of other hydrogenic defect configurations,
including some OH defects which can trap \lip \cite{Jollands2020}. There is also evidence for unidentified shallow \lip\ trap sites\cite{Jain1982,Devautour2007}.

In the simplest case, the biexponential fractions might correspond to freely diffusing and trapped \elip.
The diffusing fraction corresponds to the sharp, motionally narrowed comb resonance.
The strong hole-burning effect at late times [\Cref{hb}] indicates that
the broad line (not the sharp comb line) corresponds to the slow relaxation,
hence the diffusing fraction is the {\it fast} relaxing component at 300 K.
The decrease in this fraction $f_f$ above 150 K [see \Cref{fig:Fractions} a)],
coinciding with the known mobility onset of \lip\ in the channel, 
could then be due to diffusion to traps
(i.e., conversion from fast to slow relaxing populations).
However,
trapping is incomplete as $\sim 40$\% of the signal remains fast relaxing at 300 K.
This suggests there is a population of \elip\ in trap-free channel segments bounded by channel blockers that limit the range of
diffusion, consistent with the $\alpha$ radiotracer measurement.
The presence of the $\lambda_{\mathrm{diff}}$ peak in the slow component appears inconsistent with this interpretation, 
but it suggests the reverse process (detrapping) occurs for at least some shallow traps in this $T$ range, so that
repeated cycles of diffusion and trapping may occur.
A stochastic model of relaxation (without detrapping) is given in \Cref{trap}.
It shows that the average effect of trapping is to interpolate between the characteristic
relaxations of the free and trapped states. A more sophisticated model along these lines
(i.e., including an untrapped fraction, thermal detrapping, and possibly several trap energies) might be able to
account for the $T$ dependence above 150 K.

Thus far,
a single consistent assignment of the fractions based on the above appears untenable, as there are
substantial changes at low $T$ where \elip\ cannot be mobile.
Based on the behavior of the implanted muon\cite{Hitti1999,Baker2014}, here we consider whether formation of the neutral \elin\ could
account for the features of the data below 150 K.
This possibility is suggested by the EPR of \liz\ in quartz, and we begin with a brief summary of what is known.
Formation of \liz\ is enabled by intense Xray irradiation at low temperature\cite{Jani1986},
producing a uniform distribution of radiolytic electrons to react with the isolated \lip.
The unpaired electron yields the EPR which shows small hyperfine splittings. Electron-nucleus double resonance (ENDOR) confirms
these are from the stable $^7$Li nucleus. Detailed analysis of the EPR shows that Li is at the channel site with the unpaired electron
primarily located on the neighboring Si\cite{Weil1984,Wilson1986}. The EPR is only observed at low temperature and
its intensity decreases in two annealing steps: at 109 K, the signal is reduced by about 20\%, and above
187 K, the remaining signal is rapidly wiped out\cite{Jani1986}. Both annealing steps (but particularly the latter)
are accompanied by luminescence\cite{Halperin1986}, indicating that destruction of \liz\ is part of a charge recombination process.

In contrast to the stable \liz, the kinetics of formation (electron capture)
would play an important role for the implanted \elip\ which stops randomly, almost always isolated from other pre-existing (dilute) defects.
It can then trap an electron from its own (certainly not uniform) radiolysis track to produce \elin.
This is analogous to the formation of muonium in
quartz which depends on the implanted $\mu^+$ capturing such an electron\cite{Brewer1981,Brewer2000}.
Neutralization then depends on the distribution of radiolysis electrons, their mobility, and
competing processes such as recombination, and will, as a result, be $T$ dependent.
Neither electrons nor holes are polaronic in quartz\cite{Stoneham1997,Varley2012},
so they are quite mobile, but they will often be trapped by other defects.
The \eli\ lifetime is much longer than the $2.2\,\mu$s of the muon, so the neutralizing electron may even originate from shallow traps
rather than from radiolysis directly.
Most such traps are immobile\cite{Griscom2011,Elsayed2014},
including the self-trapped exciton (STE)\cite{Constantini2000},
and some are sufficiently stable to be useful in geochronology\cite{Preusser2009}.
However, radiolytically liberated atomic hydrogen is an electron trap that
mobilizes in the channel at $\sim 100$ K\cite{Weeks1965,Griscom1985}.
Hydrogen may thus be able to shuttle an electron to the \elip\ in a charge exchange reaction,
\begin{equation*}
\ce{H^0 + Li^+ -> H^+ + Li^0}.
\end{equation*}
We are not aware of any case where the \bnmr\ signal of paramagnetic \elin\
has been observed in a solid, so we now consider how it might manifest itself.

The hyperfine coupling to the electron spin provides a substantial effective magnetic
field at the probe nucleus. From the EPR, we expect a hyperfine coupling $A_c \sim 0.95$ MHz for \elin, so the \bnmr\
should be split (magnetically) into two ENDOR lines at $\nu_0 \pm A_c/2$.
This splitting is much larger than the frequency range in \Cref{fig:1f}.
However, it is unlikely the hyperfine field is static on the timescale of the measurement.
For stable \liz, the electron $T_{1e}$ at the EPR field (0.33 T)
is very long relative to the EPR frequency (9.3 GHz)\cite{Jani1986}. Though values are not reported,
this may be on the microsecond timescale, and $T_{1e}$ will only be longer at our much higher $B_0$.
Fluctuations of the hyperfine field at a rate comparable to the splitting would
instead produce a broad line shifted from $\nu_0$ by the thermodynamic polarization of the electron spin.

The hyperfine field would also enhance the relaxation rate.
At low field, such fast relaxation could account for the large missing fraction in \Cref{fig:Fractions} b)
which shows a clear similarity with $f_f(T)$ at high field,
but the plateau region is shifted down, suggesting field-dependent spin relaxation
controls the stepwise changes in the fractions.
It could also account for the sharp peak in the high field relaxation $\lambda_{\mathrm{pk}}$,
which appears closely related to the $T$ dependence of the fractions.
Both $\lambda_f(T)$ and $\lambda_s(T)$ are described well by the decomposition of \Cref{eq:slr-decomp}, but 
the position of $\lambda_{\mathrm{pk}}$ differs.
The slow component peak coincides with the sharp increase in $f_f$, suggesting it is due to charge and spin exchange
fluctuations of neutralization. Similarly, in low field the peaks (both slow and fast) coincide
with the sharp upward step in the missing fraction, confirming their relation to the slow component in high field.
In contrast, the fast component $\lambda_{\mathrm{pk}}$ occurs in the middle of the plateau just below the onset 
of the decrease in $f_f$. While this is substantially below the 187 K annealing of the EPR signal, it is probably connected
with annihilation of the neutral. The $\sim 2.8$ eV luminescence light associated with this annealing\cite{Martini2000,Martini2012}
is very similar to the light emitted from annihilation of the STE\cite{Trukhin1992}, suggesting one recombination pathway for the Li-trapped electron proceeds
via the STE, which may act as a long-lived (milliseconds at 100 K) transition state bottleneck due to its triplet character\cite{Hayes1984,Trukhin1992,Ismail2005}.
Magnetic field fluctuations from such an STE in the vicinity of \elip\ could provide an additional mechanism for $\lambda_{\mathrm{pk}}$.
The neutralizing electron probably also acts as a trap, inhibiting hopping of the \elin, which becomes mobile once ionized.

It is thus plausible that neutral \elin\ exists over a narrow range of temperatures between $\sim 75$ and 150 K and accounts for the 
complex $T$ dependence of the relaxation in this range. However, it would be much more compelling to have a
direct confirmation by measuring its hyperfine splitting. This will certainly be difficult and likely will require the development of new methods. In the meantime,
by repeating these measurements on a crystal with a substantially lower H content, we might confirm an essential
role for H in formation of the neutral and/or in $\lambda_{\mathrm{pk}}$.

\section{Summary \label{sec:summary}}

Having set out to study 1D diffusion of implanted \elip\ in the channels of $\alpha$-SiO$_2$, we find the situation
is more complex than expected and very different from other oxide insulators where \elip\ is immobile
(e.g., MgO\cite{MacFarlane2014-MgO}, Al$_2$O$_3$\cite{MacFarlane2023-sapphire} and ZnO\cite{ZnO2022}),
while it bears some similarities
to the 1D diffusion in rutile TiO$_2$\cite{McFadden2017}. The SLR is fast and strongly temperature dependent
with a broad diffusive peak above 200 K and a second more prominent relaxation peak at lower temperature.
The diffusive relaxation yields a mean activation barrier of \qty{178 \pm 43}{\milli\electronvolt} for isolated \lip,
in agreement with the range of other measurements and calculations.
The range over which it is the dominant relaxation mechanism limits the ability to discern characteristic features of 1D diffusion.
At 6.55 T, the \eli\ NMR spectrum has two observed components, a large amplitude broad resonance and a quadrupolar multiplet that
is only revealed by an RF comb excitation, while a third fraction has an unobserved resonance.
The quadrupole splitting is surprisingly small and increases with temperature.
Comparison of $\nu_q$ with DFT calculations confirms the Wyckoff $3a$ in-channel interstitial site for \elip.

To account for many unexpected features of the data between 75 and 150 K, it is suggested that a fraction of the implanted
ions form neutral \liz\ by trapping a radiolysis electron. Formation of the neutral should be quite general
across a wide-range of insulating materials, and the phenomenology reported here gives an indication
of how it might appear in the \bnmr\ data.
More extensive hole-burning measurements and a careful search for the hyperfine splitting of \liz\ would help to test this
hypothesis. Depositing thin electrodes on the crystal to modulate the neutralization process via application of an electric field would also help to confirm
the neutral. With some knowledge of its binding energy, spectroscopic confirmation might be obtained by illuminating the sample during the \bnmr\ measurement to
alter its electronic state directly.

\begin{acknowledgments}

We are grateful to R.\ Abasalti, D.\ Arseneau, B.\ Hitti, B.\ Smith, and D.\ Vyas for technical assistance.
This work was supported by NSERC Discovery grants, NSERC CREATE IsoSiM fellowships (A.C.\ and R.M.L.M.),
and SBQMI QuEST fellowships (D.F., V.L.K., and J.O.T.).
This work was, in part, supported by funding from the Max Planck-UBC-UTokyo Center
for Quantum Materials and the Canada First Research Excellence Fund, Quantum Materials and Future Technologies Program.
We thank D.\ Ceresoli, J.\ Adelman, K.\ Foyevtsova, P.\ Macau, and I.\ Elfimov for help with the DFT calculations.
Structural figures were made with CrystalMaker.
Additionally, we thank Y.\ Qi and S.-Y.\ Kim for some clarification of Ref.\ \onlinecite{Kim2014}.
W.A.M. acknowledges the hospitality of LNCMI Grenoble for part of this work.
\end{acknowledgments}

\appendix

\section{Sample Comparison \label{sampdep}}

\begin{figure}%[tbh]
\centering
\includegraphics[width=\columnwidth]{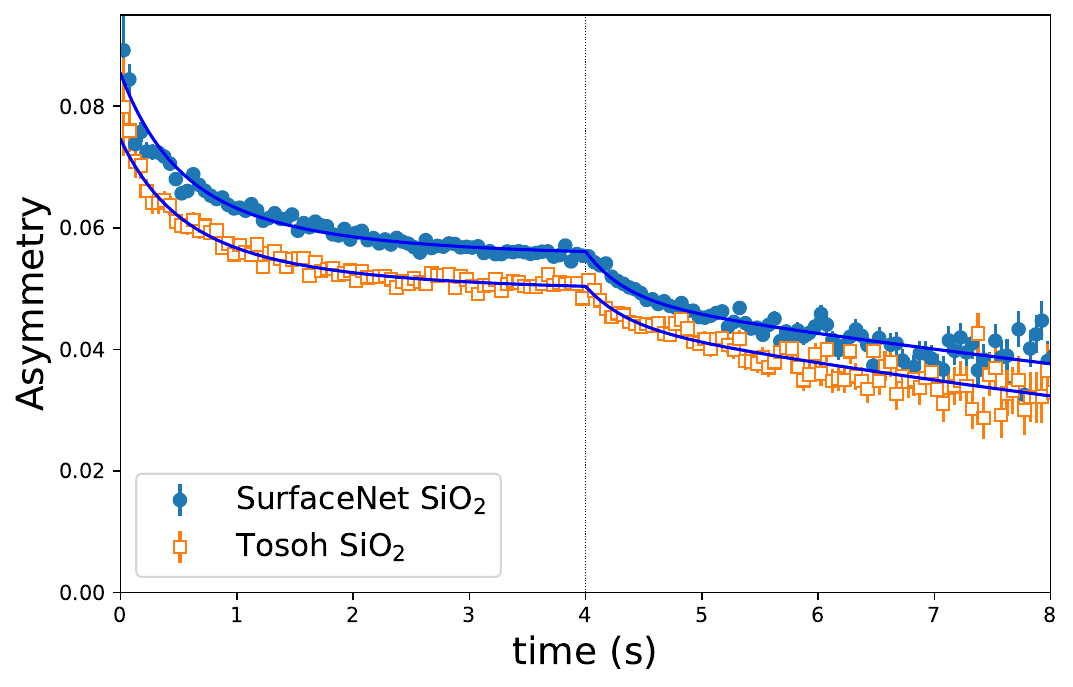}
\caption{ Spin relaxation of \elip\ implanted into two different samples of \ce{SiO2} at room temperature and 6.55 T.
The similarity between the samples confirms that the relaxation is typical of high purity quartz.}
\label{fig:samp-comp}
\end{figure}

At high field, we compare the relaxation data in the two quartz samples from different sources at 6.55 T and 300 K , see 
\Cref{fig:samp-comp}. The data are not artificially offset, and the difference in amplitude is within the
expected range of variability as the data were not taken in the same year.
The similarity of the data indicates that the relaxation in the crystal presented in the main text is not
sample specific, but appears characteristic of high purity quartz.

\section{Field Dependence of the Relaxation at 300~K \label{bdep}}

\begin{figure}%[tbh]
\centering
\includegraphics[width=\columnwidth]{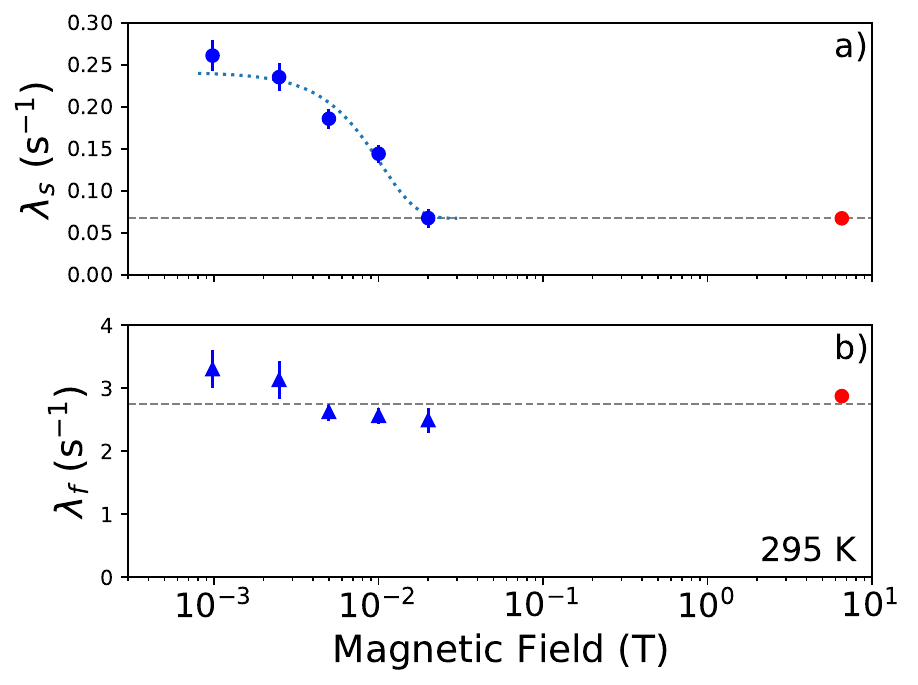}
\caption{ Rates from biexponential fits to the relaxation as a function of applied field at 295 K with a shared fraction $f_s$.
At the lowest fields, the amplitude is further reduced. The red points are the high field data where there is no missing fraction.
The horizontal lines show the rates at 20 mT are similar to those at 6.55 T, though the amplitude is significantly reduced.}
\label{fig:ratesvfield}
\end{figure}

At 295 K, the relaxation was measured as a function of $B_0$ in the low field regime.
The data can be fit (reduced $\chi^2 = 1.1$) with a biexponential with a shared fraction $f_f = 0.704(8)$.
The resulting rates (shown in \Cref{fig:ratesvfield}) demonstrate that the remnant signal has relaxation rates at 20 mT
comparable to those at much higher field, despite the much smaller overall amplitude. The horizontal dashed lines are the
average of the high field and 20 mT values. Here the relaxation is largely $\lambda_{\mathrm{bg}}$
which is probably magnetic rather than quadrupolar in origin.
Below 20 mT, the rate of the slow component climbs,
while the fast component remains relatively field independent. The similarity of rates suggests that (at room temperature)
the same mechanism is predominant above about 20 mT, while at lower field an additional channel appears to dominate for the slow component.

\section{Density Functional Theory Calculations \label{dft}}

\begin{figure}%[tbh]
\centering
\includegraphics[width=\columnwidth]{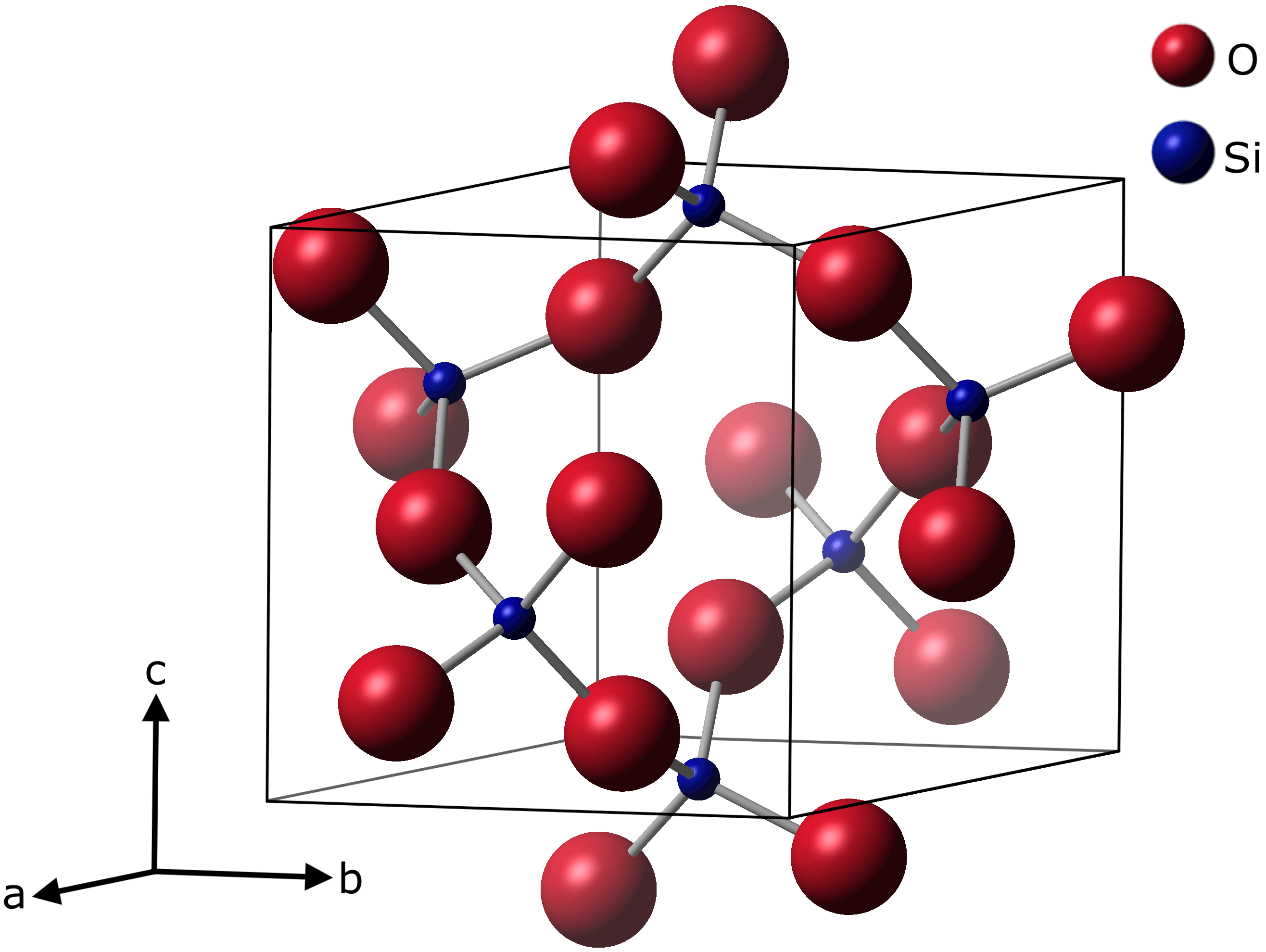}
\caption{The conventional trapezoidal unit cell of \quartz\ (space group 152/154 depending on handedness) containing 3 formula units
shown here for the Si atom in the basal plane at fractional coordinates $(0.53,0.53,0.0)$.
The central axis of the channel in \Cref{fig:channel} corresponds to the verical ($c$) unit cell edges.
The room temperature lattice constants are  $a = b =  4.91239(4)$ \AA\ and $c = 5.40385(7)$ \AA \cite{Will1988}, with angles
$\alpha = \beta = 90^\circ$ and $\gamma = 120^\circ$.
}
\label{fig:unitcell}
\end{figure}

\begin{figure}%[tbh]
\centering
\includegraphics[width=\columnwidth]{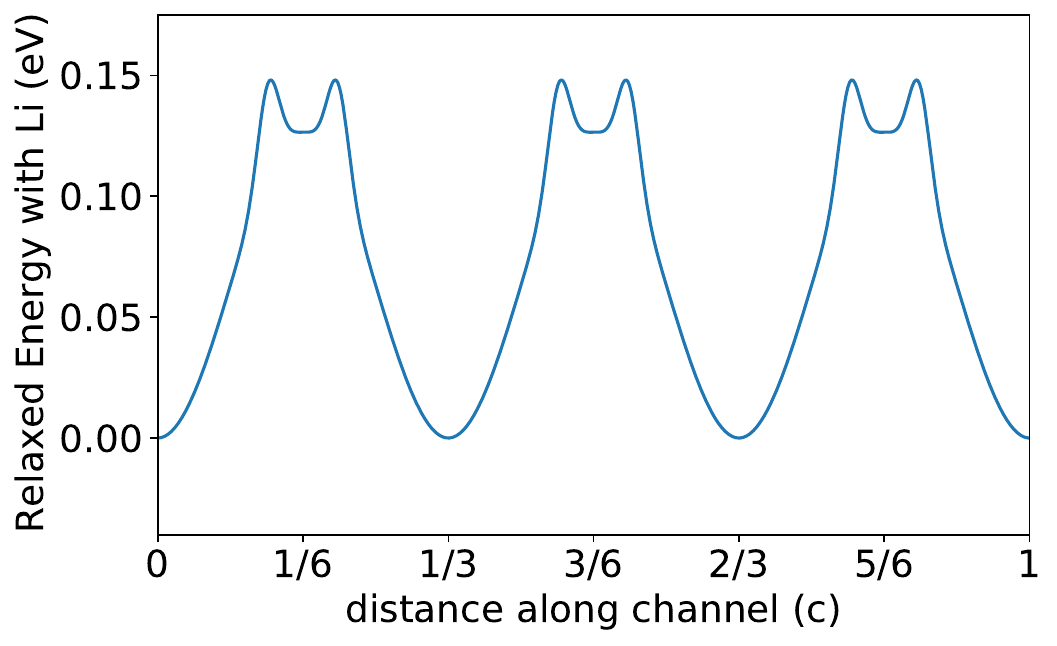}
\caption{Schematic potential barrier for dilute \lip\ along the channel in $\alpha$-SiO$_2$ based on the
two relaxed site energies in a $3\times3\times4$ supercell.}
\label{fig:barriercartoon}
\end{figure}

For the DFT calculations, we used {\sc quantum espresso} \cite{QE-2020,QE-2017,QE-2009} with
Troullier-Martins\cite{Troullier1991} norm-conserving pseudopotentials optimized for solid state NMR calculations\cite{Tantardini2022}.
We constructed supercells based on the conventional unit cell shown in \Cref{fig:unitcell}. 
To obtain a calculated EFG comparable to experiment, we fixed the lattice constants at the experimental values
and allowed the system to relax with a constant unit cell volume.
The kinetic energy cutoff for wavefunctions was $80$ Ry, while for the charge density and potential $320$ Ry was used.
The energetic convergence threshold for self-consistency was set to $10^{-10}$ Ry and the mixing factor (beta) was $0.5$.
We modeled the Li$^+$ in two ways: the ion with a uniform compensating (jellium) charge and a neutral Li atom.
Both gave similar results as expected\cite{Bonfa2015}, and only the former are reported here. 
We made no effort to try to model the metastable neutral Li$^0$.
To find the lowest energy sites, a sampling of initial Li coordinates were run with a smaller $2\times2\times2$ supercell (73 atoms)
and less stringent convergence criteria.
For the final energies and EFG, the calculation used a $4\times4\times4$ reciprocal space grid with no offset\cite{Monkhorst1976}
and a $3\times3\times4$ supercell (325 atoms) to approximate the dilute limit (Li at $\sim 9300$ ppm relative to Si).
The EFG calculation used the {\sc gipaw} implementation\cite{Varini2013} packaged with {\sc quantum espresso}.
The extent of the lattice relaxation around the interstitial Li was quantified by the magnitude of the displacement
vector from relaxed supercells with and without \lip. Only the first few near-neighbour shells showed appreciable relaxation
with all displacements less than 0.16 \AA.

\section{Hole Burning \label{hb}}

\begin{figure}%[tbh]
\centering
\includegraphics[width=\columnwidth]{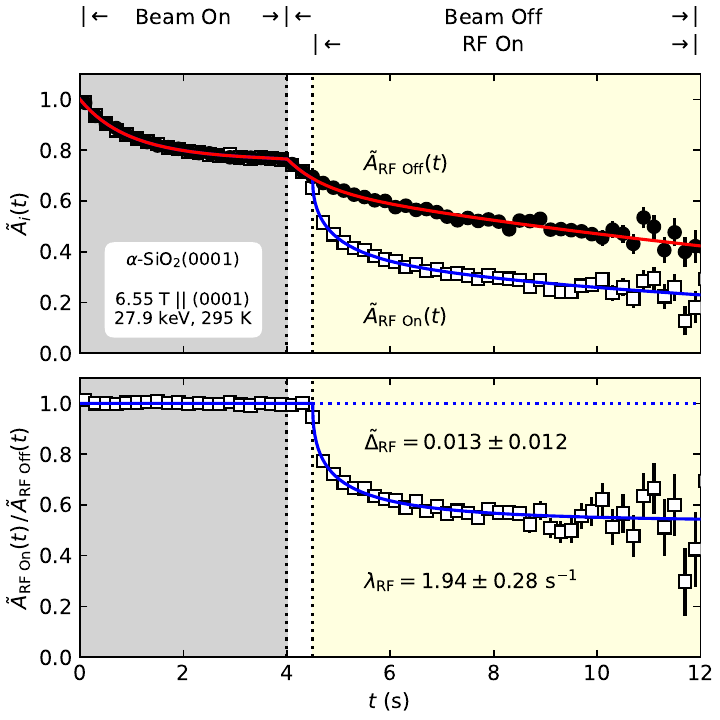}
\caption{Hole-burning measurement at 6.55 T and 295 K.
The single tone RF excitation at the centre of the
resonance is turned on 0.5 s after the end of the beam pulse.
}
\label{fig:hb}
\end{figure}

\begin{figure}%[tbh]
\centering
\includegraphics[width=\columnwidth]{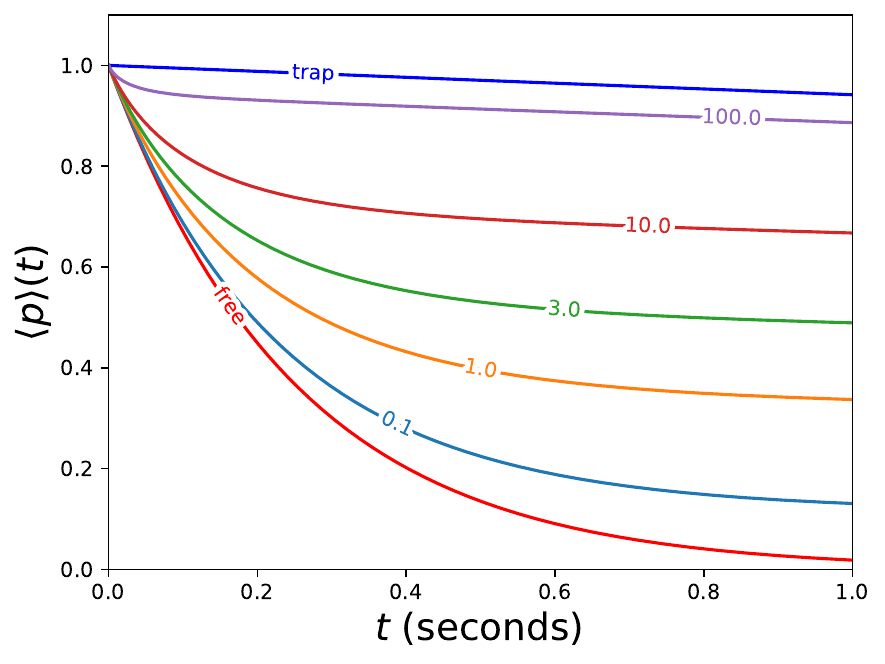}
\caption{The average spin relaxation for a stochastic model of
1D diffusion and trapping with trap concentration 2500 ppm.
The diffusing fraction is assumed to relax rapidly while the trapped fraction is slow.
The average polarization from \Cref{eq:avpol} is labelled by the hop rate $W$ in kHz and
shows a characteristic interpolation between the free and trapped relaxation functions. 
No detrapping is included.
}
\label{fig:trapping}
\end{figure}

A hole-burning experiment was done at 295 K and 6.55 T.
With pulsed beam, the RF at the center of the broad line in \Cref{fig:1f}
is switched on at a specific time relative to the beam pulse. 
Any \eli\ resonant with the RF will be depolarized in ms, leading
to a sharp downward step $\Delta_{\mathrm{RF}}$ in the time differential asymmetry
with a 10 ms time-resolution. Much slower spectral diffusion brings other \eli\ onto resonance over time
leading to an apparent relaxation much faster than the SLR rate seen in the absence of RF.
When the stochastic spectral diffusion of the resonance frequency occurs in many small steps, the 
resulting hole-burning relaxation is often of a stretched exponential form
$\propto \exp ( - \sqrt{ \lambda_{\mathrm{RF}} - t_{\mathrm{RF}}  } )$
The data and fits together with a calibration run without RF are shown in \Cref{fig:hb}.
The spectral diffusion relaxation is clearly a strong effect and will amplify
the corresponding time-integral resonance in \Cref{fig:1f} by at least a factor of 10.
The spectral diffusion relaxation also confirms that there are slow fluctuations of the environment of
a significant fraction of the \eli\ in addition to the fast fluctuations that give rise to the diffusive SLR.

\section{Diffusion and Trapping in 1D \label{trap}}

Here we illustrate the behavior of the depolarization function from a stochastic model of trapping in 1D.
Defining $f(t)$ as the freely diffusing fraction at time $t$, the ``survival fraction'' in the argot of survival analysis.
Let $\eta(t) = -df/dt$ be the probability per unit time that a diffuser is trapped in time interval $(t,t+dt)$.
Movaghar {\it et al.}\cite{Movaghar1982} considered 1D diffusion and trapping in some detail.
Here we use their expression for the `First Passage Time' estimate for $f$ in 1D, namely,
\begin{align*}
f_{\mathrm{FPT}} = &  f_0 \left( 1-\int_0^t d\tau \, \eta(\tau) \right) \\
                 = & f_0 \exp\left( -x \int_0^t d\tau \, 2W e^{-2W\tau}[I_0(2W\tau) + I_1(2W\tau)] \right),
\end{align*}
where $I_{0,1}$ are modified Bessel functions, $W$ is the hop rate
and $x$ is the relative concentration of the trap sites.
Next we assume the trapped and free ions have characteristic
single exponential relaxation functions, $p_{\mathrm{free}}$ and $p_{\mathrm{trap}}$,
with rates $\lambda_{\mathrm{free}} = 4$ and $\lambda_{\mathrm{trap}} = 0.06$ s$^{-1}$
based on the 6.55 T data at about 200 K, and then we apply the stochastic trapping model\cite{Herlach1981} to 
obtain the average relaxation function $\langle p \rangle$,
\begin{equation}
\langle p \rangle = f(t) p_{\mathrm{free}}(t) + \int_0^td\tau \, \eta(\tau) p_{\mathrm{free}}(\tau)p_{\mathrm{trap}}(t-\tau) .
\label{eq:avpol}
\end{equation}
Assuming a negligible initial trapped fraction $f_0=1$ and a trap concentration appropriate to hydrogen,
numeric integration gives the results shown in \Cref{fig:trapping}. The average
$\langle p \rangle$ interpolates between $p_{\mathrm{free}}$ and $p_{\mathrm{trap}}$ for
relatively low hop rates at this concentration of traps. Qualitiatively, it resembles
a biexponential where, as the hop rate increases, the fast fraction diminishes in amplitude
and slows in rate before increasing in rate again.

\bibliography{quartz-references.bib}

\end{document}

%% file: preamble/packages.tex
\usepackage[T1]{fontenc} 
\usepackage[utf8]{inputenc}
\usepackage{newtxtext}
\usepackage{newtxmath}
\usepackage{graphicx}
\usepackage{siunitx} 
\usepackage[version=4]{mhchem}
\usepackage{booktabs}
\usepackage[english]{babel}
\usepackage{csquotes}
\usepackage[unicode,colorlinks=true,allcolors=blue]{hyperref}
\usepackage{cleveref}
\usepackage{microtype}

%% file: preamble/abbreviations.tex
\newcommand{\bnmr}{$\beta$-NMR}

\newcommand{\slr}{SLR}

\newcommand{\rf}{RF}

\newcommand{\eli}{\textsuperscript{8}{Li}}

\newcommand{\elip}{\textsuperscript{8}{Li}\textsuperscript{+}}
\newcommand{\elin}{\textsuperscript{8}{Li}\textsuperscript{0}}
\newcommand{\lip}{{Li}\textsuperscript{+}}
\newcommand{\liz}{{Li}\textsuperscript{0}}

\newcommand{\quartz}{$\alpha$-\ce{SiO2}}